
\input harvmac
\parskip 7pt plus 1pt
\noblackbox

\def\la{\mathrel{\mathpalette\fun <}}
\def\ga{\mathrel{\mathpalette\fun >}}
\def\fun#1#2{\lower3.6pt\vbox{\baselineskip0pt\lineskip.9pt
  \ialign{$\mathsurround=0pt#1\hfil##\hfil$\crcr#2\crcr\sim\crcr}}}
\relax


\lref\BP{R. Bousso and J. Polchinski, ``Quantization of Four-form
Fluxes and Dynamical Neutralization of the Cosmological
Constant,'' JHEP {\bf 0006} (2000) 006, hep-th/0004134.}

\lref\MSS{A. Maloney, E. Silverstein and A. Strominger, ``de Sitter
Space in Non-Critical String Theory,'' hep-th/0205316.}

\lref\linde{A.D. Linde, {\it Particle Physics and Inflationary
Cosmology}, Harwood, Chur, Switzerland (1990).}

\lref\liddle{A.R. Liddle and D.H. Lyth, {\it Cosmological
Inflation and Large-Scale Structure}, Cambridge University Press,
Cambridge, England (2000).}

\lref\tyeangle{N. Jones, H. Stoica and S.H. Tye, ``Brane
Interaction as the Origin of Inflation,'' JHEP {\bf 0207} (2002) 051,
hep-th/0203163.}

\lref\shafi{B. Kyae and Q. Shafi, ``Branes and Inflationary Cosmology,''
Phys. Lett. {\bf B526} (2002) 379, hep-ph/0111101.}

\lref\ganor{O. Ganor, ``A Note on Zeroes of Superpotentials in
F-theory,'' Nucl. Phys. {\bf B499} (1997) 55, hep-th/9612077.}

\lref\bbhl{K. Becker, M. Becker, M. Haack and J. Louis,
``Supersymmetry Breaking and Alpha-Prime Corrections to Flux
Induced Potentials,'' JHEP {\bf 0206} (2002) 060, hep-th/0204254.}

\lref\dvalitye{G. Dvali and S.H. Tye, ``Brane Inflation,'' Phys.
Lett. {\bf B450} (1999) 72, hep-th/9812483.}

\lref\juanAdS{J. Maldacena, ``The Large $N$ Limit of
Superconformal Field Theories and Supergravity,'' Adv. Th. Math.
Phys. {\bf 2} (1998) 231, hep-th/9711200.}

\lref\hhk{C. Herdeiro, S. Hirano and R. Kallosh, ``String Theory
and Hybrid Inflation/Acceleration,'' JHEP {\bf 0112} (2001) 027,
hep-th/0110271.}

\lref\others{ S. Alexander, ``Inflation from D - Anti-D-Brane
Annihilation,'' Phys. Rev. {\bf D65} (2002) 023507,
hep-th/0105032\semi G. Dvali, Q. Shafi and S. Solganik, ``D-brane
Inflation,'' hep-th/0105203\semi C.P. Burgess, M. Majumdar, D.
Nolte, F. Quevedo, G. Rajesh and R.J. Zhang, ``The Inflationary
Brane-Antibrane Universe,'' JHEP {\bf 07} (2001) 047,
hep-th/0105204\semi G. Shiu and S.H. Tye, ``Some Aspects of Brane
Inflation,'' Phys. Lett. {\bf B516} (2001) 421, hep-th/0106274 \semi D. Choudhury, D. Ghoshal, D.P. Jatkar,
S. Panda, ``Hybrid Inflation and Brane-Antibrane System,'' hep-th/0305104.}

\lref\quevedo{F. Quevedo, ``Lectures on String/Brane Cosmology,''
hep-th/0210292.}

\lref\GKP{S. Giddings, S. Kachru and J. Polchinski, ``Hierarchies
from Fluxes in String Compactifications,'' Phys. Rev. {\bf D66}
(2002) 106006, hep-th/0105097.}

\lref\dWG{O. DeWolfe and S. Giddings, ``Scales and Hierarchies in
Warped Compactifications and Brane Worlds,'' Phys. Rev. {\bf D67}
(2003) 066008, hep-th/0208123.}

\lref\Giddings{S. Giddings, ``The Fate of Four Dimensions,'' Phys.
Rev. {\bf D68} (2003) 026006, hep-th/0303031.}

\lref\KKLT{S. Kachru, R. Kallosh, A. Linde and S. P.  Trivedi, ``de
Sitter Vacua in String Theory,'' hep-th/0301240, to appear in
Phys. Rev. {\bf D}.}

\lref\KPV{S. Kachru, J. Pearson and H. Verlinde, ``Brane/Flux
Annihilation and the String Dual of a Nonsupersymmetric Field
Theory,'' JHEP {\bf 0206} (2002) 021, hep-th/0112197.}

\lref\KofLinde{L. Kofman and A. Linde, ``Problems with Tachyon
Inflation,'' JHEP {\bf 0207} (2002) 004, hep-th/0205121.}

\lref\banksdine{T. Banks and M. Dine, ``Coping with Strongly
Coupled String Theory,'' hep-th/9406132.}

\lref\kalletal{R. Kallosh, work in progress.}

\lref\RS{L. Randall and R. Sundrum, ``A Large Mass Hierarchy from
a Small Extra Dimension,'' Phys. Rev. Lett. {\bf 83} (1999) 3370,
hep-ph/9905221.}

\lref\GW{W. Goldberger and M. Wise, ``Modulus Stabilization with
Bulk Fields,'' Phys. Rev. Lett. {\bf 83} (1999) 4922,
hep-ph/9907447.}

\lref\KS{I. Klebanov and M.J. Strassler, ``Supergravity and a
Confining Gauge Theory:  Duality Cascades and $\chi$SB Resolution
of Naked Singularities,'' JHEP {\bf 0008} (2000) 052,
hep-th/0007191.}

\lref\FiscSuss{
S. Hellerman, N. Kaloper and L. Susskind, ``String Theory and
Quintessence,'' JHEP {\bf 0106} (2001) 003, hep-th/0104180\semi
W. Fischler, A. Kashani-Poor, R. McNees and S.
Paban, ``The Acceleration of the Universe: A Challenge for String
Theory,'' JHEP {\bf 0107} (2001) 003, hep-th/0104181.}

\lref\Kallosh{K. Dasgupta, C. Herdeiro, S. Hirano and R. Kallosh,
``D3/D7 Inflationary Model and M-theory,'' Phys. Rev. {\bf D65}
(2002) 126002, hep-th/0203019.}

\lref\Garcia{J. Garcia-Bellido, R. Rabadan and F. Zamora,
``Inflationary Scenarios from Branes at Angles,'' JHEP {\bf 0201}
(2002) 036, hep-th/0112147.}

\lref\BinetruyXJ{ P.~Binetruy and G.~R.~Dvali, ``D-term
inflation,'' Phys.\ Lett.\ B {\bf 388} (1996) 241, hep-ph/9606342.
}

\lref\wittenred{ E.~Witten, ``Dimensional Reduction Of Superstring
Models,'' Phys.\ Lett.\ B {\bf 155} (1985) 151.

}

\lref\andreian{A. Linde, ``Inflation, quantum cosmology and
the anthropic principle,'' hep-th/0211048.}

\lref\lenny{L. Susskind, ``The Anthropic Landscape of String
Theory,'' hep-th/0302219.}

\lref\shanta{R. Brustein, S. De Alwis and E. Novak, ``Inflationary
Cosmology in the Central Region of String/M-theory Moduli space,''
Phys. Rev. {\bf D68} (2003) 023517, hep-th/0205042\semi
R. Brustein, S. De Alwis and E. Novak, ``M-theory Moduli Space
and Cosmology,'' Phys. Rev. {\bf D68} (2003) 043507, hep-th/0212344.}

\lref\Douglas{S. Ashok and M. Douglas, ``Counting Flux Vacua,''
hep-th/0307049\semi
M. Douglas, ``The statistics of string/M theory vacua,''
hep-th/0303194.
}

\lref\SeibergWitten{ N.~Seiberg and E.~Witten, ``The D1/D5 System
and Singular CFT,'' JHEP {\bf 9904} (1999) 017, hep-th/9903224.

}


\lref\Lindefast{ A.~Linde, ``Fast-roll Inflation,'' JHEP {\bf
0111} (2001) 052, hep-th/0110195. }


\lref\KofLinde{L. Kofman and A. Linde, ``Problems with Tachyon Inflation,''
JHEP {\bf 0207} (2002) 004, hep-th/0205121.}

\lref\Sen{A. Sen, ``Tachyon Condensation on the Brane Anti-Brane System,''
JHEP {\bf 9808} (1998) 012, hep-th/9805170.}

\lref\tachac{M.R. Garousi, ``Tachyon Couplings on Non-BPS D-branes
and Dirac-Born-Infeld Action,'' Nucl. Phys. {\bf B584} (2000) 584,
hep-th/0003122\semi E.A. Bergshoeff, M. de Roo, T.C. de Wit, E.
Eyras and S. Panda, ``T-Duality and Actions for Non-BPS D-branes,"
JHEP {\bf 0005} (2000) 009, hep-th/0003221\semi J. Kluson,
``Proposal for Non-BPS D-brane Action,'' Phys. Rev. {\bf D62}
(2000) 126003, hep-th/0004106.}

\lref\senvertex{A.~Sen, ``Dirac-Born-Infeld Action on the Tachyon
Kink and Vortex,'' hep-th/0303057.
}

\lref\liddlenew{P.~P.~Avelino and A.~R.~Liddle, ``Cosmological
Perturbations and the Reionization Epoch,'' astro-ph/0305357.
}

\lref\vilet{ A.~Vilenkin, ``The Birth of Inflationary Universes,''
Phys.\ Rev.\ D {\bf 27} (1983) 2848.
}

\lref\tyecosmic{
S.~Sarangi and S.~H.~Tye,
``Cosmic String Production Towards the End of Brane Inflation,''
Phys.\ Lett.\ B {\bf 536} (2002) 185,
hep-th/0204074;
N.~T.~Jones, H.~Stoica and S.~H.~Tye,
``The Production, Spectrum and Evolution of Cosmic Strings in
Brane Inflation,''
Phys.\ Lett.\ B {\bf 563} (2003) 6,
hep-th/0303269;
L.~Pogosian, S.~H.~Tye, I.~Wasserman and M.~Wyman,
``Observational Constraints on Cosmic String Production During Brane
Inflation,''
Phys.\ Rev.\ D {\bf 68} (2003) 023506,
hep-th/0304188.
}

\lref\linet{ A.~D.~Linde, ``Eternally Existing Selfreproducing
Chaotic Inflationary Universe,'' Phys.\ Lett.\ B {\bf 175} (1986)
395.
}

\lref\lindeXX{ A.~D.~Linde, D.~A.~Linde and A.~Mezhlumian, ``From
the Big Bang Theory to the Theory of a Stationary Universe,''
Phys.\ Rev.\ D {\bf 49} (1994) 1783, gr-qc/9306035.
}

\lref\BennettBZ{ C.~L.~Bennett {\it et al.}, ``First Year
Wilkinson Microwave Anisotropy Probe (WMAP) Observations:
Preliminary Maps and Basic Results,'' astro-ph/0302207.
}

\lref\hybrid{A.~D.~Linde, ``Axions in Inflationary Cosmology,''
Phys.\ Lett.\ B {\bf 259} (1991) 38;
A.~D.~Linde, ``Hybrid Inflation,'' Phys.\ Rev.\ D {\bf 49} (1994)
748, astro-ph/9307002.
}

\lref\fterm{E.~J.~Copeland, A.~R.~Liddle, D.~H.~Lyth,
E.~D.~Stewart and D.~Wands, ``False Vacuum Inflation with Einstein
Gravity,'' Phys.\ Rev.\ D {\bf 49} (1994) 6410, astro-ph/9401011;
G.~R.~Dvali, Q.~Shafi and R.~Schaefer, ``Large Scale Structure and
Supersymmetric Inflation without Fine Tuning,'' Phys.\ Rev.\
Lett.\ {\bf 73} (1994) 1886, hep-ph/9406319; A.~D.~Linde and
A.~Riotto, ``Hybrid Inflation in Supergravity,'' Phys.\ Rev.\ D
{\bf 56} (1997) 1841, hep-ph/9703209.
}

\lref\cyclic{ P.~J.~Steinhardt and N.~Turok, ``Cosmic Evolution in
a Cyclic Universe,'' Phys.\ Rev.\ D {\bf 65} (2002) 126003,
hep-th/0111098.
}

\lref\lindehawking{ A.~Linde, ``Inflationary Theory versus
Ekpyrotic / Cyclic Scenario,'' hep-th/0205259.
}

\lref\frey{A. Frey, M. Lippert and B. Williams, ``The Fall of
Stringy de Sitter,'' hep-th/0305018.}

\lref\quevedoDS{C. Escoda, M. Gomez-Reino and F. Quevedo,
``Saltatory de Sitter String Vacua,''
hep-th/0307160.}

\lref\ContaldiHI{ C.~R.~Contaldi, H.~Hoekstra and A.~Lewis,
``Joint CMB and Weak Lensing Analysis: Physically Motivated
Constraints on Cosmological Parameters,'' Phys.\ Rev.\ Lett.\ {\bf
90} (2003) 221303, astro-ph/0302435.
}

\hskip 1cm
\vskip -1 cm

\Title{\vbox{\baselineskip12pt \hbox{hep-th/0308055}
\hbox{SLAC-PUB-9669} \hbox{SU-ITP-03/18} \hbox{TIFR/TH/03-06} }}
{\vbox{\centerline{Towards Inflation in String Theory}}}

\centerline{Shamit
Kachru,$^{a,b}$\footnote{}{skachru@stanford.edu,
kallosh@stanford.edu, alinde@stanford.edu,} Renata Kallosh,$^{a}$
Andrei Linde,$^{a}$
}
\centerline{
 Juan Maldacena,$^{c}$ \footnote{}{malda@ias.edu,
 lpm@itp.stanford.edu, sandip@tifr.res.in } Liam
McAllister,$^{a}$
and
Sandip P. Trivedi$^{d}$
}

\bigskip\centerline{$^a$ \it Department of Physics, Stanford
University, Stanford, CA 94305, USA} \smallskip \centerline{$^b$
\it SLAC, Stanford University,  Stanford, CA 94309, USA}
\smallskip \centerline{$^c$ \it Institute for Advanced Study,
Princeton, NJ 08540,
 USA}
\smallskip \centerline{$^d$ \it Tata Institute of Fundamental
Research, Homi Bhabha Road, Mumbai 400 005, INDIA}

\vskip .2in

We investigate the embedding of brane inflation into stable
compactifications of string theory.  At first sight a warped
compactification geometry seems to produce a naturally flat
inflaton potential, evading one well-known difficulty of
brane-antibrane scenarios.  Careful consideration of the closed
string moduli reveals a further obstacle: superpotential
stabilization of the compactification volume typically modifies
the inflaton potential and renders it too steep for inflation.  We
discuss the non-generic conditions under which this problem does
not arise. We conclude that brane inflation models can only work
if restrictive assumptions about the method of volume
stabilization, the warping of the internal space, and the source
of inflationary energy are satisfied.  We argue that this may not
be a real problem, given the large range of available
fluxes and background geometries in string theory.

\eject


\newsec{Introduction}

Inflation provides a compelling explanation for the homogeneity
and isotropy of the universe and for the observed spectrum of
density perturbations \refs{\linde,\liddle}. For this reason, we
would hope for inflation to emerge naturally from any fundamental
theory of microphysics.  String theory is a promising candidate
for a fundamental theory, but there are significant obstacles to
deriving convincing models of inflation from string theory.

 One problem is that string compactifications come with moduli
fields which control the shape and size of the compactification
manifold as well as the string coupling.  Inflation is possible
only if these fields are either stable or else have relatively
flat potentials which do not cause fast, non-inflationary rolling
in field space.  Controlling the moduli in this way is a difficult
problem.  In particular, the potential for the dilaton and for the
compactification volume tends to be a rather steep function
\FiscSuss.

A second problem is that the inflaton potential itself must be
exceptionally flat to ensure prolonged slow-roll inflation.  A
successful microphysical theory would naturally produce such a
flat potential. Since the flatness condition for the potential
involves the Planck scale one should ensure that quantum gravity
corrections do not spoil it.  Hence, the problem should be
analyzed in a theory of quantum gravity, such as string theory.
The hope of brane-antibrane inflation scenarios is that the
brane-antibrane interaction potential can play the role of the
inflaton potential (see \quevedo\ for a nice review), but it is
well known that this potential is {\it{not}} naturally flat. Since
in string theory one cannot fine-tune by hand, but only by varying
background data (like the compactification manifold or the choice
of flux), one concludes that in generic compactifications, brane
inflation will not work.  However, the many choices of flux and
compactification make possible a considerable degree of discrete
fine-tuning, so for very special choices of the background one
would expect to find potentials which are sufficiently flat for
inflation.

In this note we discuss these problems in the concrete context of
the warped type IIB compactifications described in e.g.
\refs{\GKP,\dWG}.  One reason for working in this setting is that
one can sometimes stabilize all the moduli in a geometry of this
type, avoiding the first problem mentioned above.  In addition,
the constructions of \GKP\ naturally admit D3-branes and
anti-D3-branes transverse to the six compact dimensions.
Furthermore, one could wish for a model which accommodates both
inflation and the present-day cosmic acceleration.  This might be
possible if one could construct inflationary models which
asymptote at late times to the de Sitter vacua of \KKLT\ (or
variants on that construction, as described in e.g. \refs{\frey,
\quevedoDS}; earlier constructions in non-critical string
theory appeared in \MSS).  As these
vacua included one or more anti-D3-branes in a warped type IIB
background, it is quite natural to consider brane-antibrane
inflation in this context.

Our idea, then, is to begin with the de Sitter vacua constructed
in \KKLT, add a mobile D3-brane, and determine whether the
resulting potential is suitable for inflation.
\footline={\hss\tenrm\folio\hss}
For the impatient reader, we summarize our findings here.
We find that modest warping of the compactification geometry
produces an extremely flat brane-antibrane interaction potential,
provided that we neglect moduli stabilization. This solves the
second problem listed above.  However, a new problem
appears when we incorporate those terms in the potential which led,
in the construction of \KKLT,  to the stabilization of the volume
modulus. We show that generic volume-stabilizing superpotentials
also impart an unacceptably large mass
to the inflaton,
halting inflation.

While these conclusions are ``generic,'' it is very important to
emphasize that the problem of the inflaton mass might be
circumvented in at least two different ways.  First, the
stabilization mechanism for the moduli might be different from
that in \KKLT. For example, the volume modulus could be stabilized
by corrections to the K\"ahler potential, which, as we will see,
can naturally circumvent this problem.
 Second, the mobile brane might be located not
at a generic point in the compact manifold but close to some
preferred point. If the location of the D3-brane is appropriately
chosen then there could be significant corrections to the
superpotential.  In general models, the superpotential may be a
rather complicated function of both the brane positions and the
volume modulus.  Little is known about the form of these
nonperturbative superpotentials in string compactifications.  Our
arguments show that if the functional form of the superpotential
is generic then inflation does not occur.  Nevertheless, it seems
quite likely, given the range of available fluxes and background
geometries, that cases exist which are sufficiently non-generic to
permit inflation, although with predictions which are altered from
those of naive brane inflation.\foot{This point is made more
quantitative in Appendix F, where we explain that the degree of
non-genericity required corresponds roughly to a fine-tune of one
part in 100.}

Our conclusions should be viewed as a first pass through the class
of brane inflation models, in the context of the moduli
stabilization mechanism which has recently been developed in
\refs{\GKP, \KKLT}.  Once the non-perturbative superpotentials
involved in such constructions are better understood, and/or as
soon as other mechanisms for moduli stabilization become
available, one could re-examine brane inflation in light of this
further concrete knowledge.  This may well lead to a precise
determination of the non-generic cases where working models of
brane inflation in string theory can be realized.

Our analysis clearly indicates that any viable inflation scenario in
string theory has to address the moduli stabilization problem.
Since essentially all papers on the subject, to the best of our knowledge,
have ignored the problem of moduli stabilization, their
conclusions are questionable in view of our results. In
particular, should a more detailed analysis reveal the possibility
of inflation in various non-generic situations, as suggested
above, we expect that the resulting inflationary parameters will
typically be quite different from those calculated in the existing
literature by neglecting moduli stabilization.

This paper is organized as follows. In \S2 we review basic facts
about brane-antibrane inflation \refs{\dvalitye,\others}, with
special attention to the case of D3-branes, and discuss some
generic problems for such models.  In \S3 we show that warping of
the geometry can help with some of these problems.  In \S4 we
explain one method of embedding the warped inflation scenario into
string theory, using the warped compactifications of \GKP.  In
\S5 we describe further problems that arise in the string theory
constructions when one tries to stabilize the overall volume
modulus. Generic methods of stabilization (e.g. via a
nonperturbative superpotential) modify the inflaton potential and
make inflation difficult to achieve.  We discuss several ways to
overcome this problem.
We conclude with some general remarks in \S6.

Appendix A contains a general discussion of the gravitational
interaction of an (unwarped) brane-antibrane pair, and
demonstrates that the potentials which arise are
typically not flat enough to lead to prolonged inflation. In Appendix B we
specialize to a warped background and derive the interaction
potential.  In Appendix C we explore the detailed properties of
inflation in warped brane-antibrane models, assuming that a
solution to the challenges of \S5 has been found.  In Appendix D
we explain that eternal inflation may be possible in this
scenario. In Appendix E we discuss the exit from inflation and
point out that the production of undesirable metric perturbations
due to cosmic strings, which are typically created during
brane-antibrane annihilation, is highly suppressed in warped
models. Finally, in Appendix F we discuss the possibility of
fine-tuning of the inflaton potential in order to achieve an
inflationary regime.

After completing this work, we became aware of the papers \shanta, in which
related issues are addressed.

\newsec{Brief Review of $D3 / {\overline{D3}}$ Inflation}

In brane-antibrane inflation one studies the relative motion of a
brane and an antibrane which are initially separated by a distance
$r$ on the compactification manifold $M$. One should assume $r \gg
l_{s}$, so that the force is well approximated by the Coulomb
attraction due to gravity and RR fields.  Then the potential takes
the form \eqn\bantibpot{V(r) = 2T_{3}\left( 1 - { 1 \over 2 \pi^3}
{ T_{3} \over M_{10,Pl}^8 r^4} \right)~.} where $M_{10,Pl}$ is the
ten-dimensional Planck scale, defined by $ 8 \pi G_{10,N} =
M_{10,Pl}^{-8}$, and $T_3$ is the tension of a D3-brane.
 In terms of a canonically normalized
scalar field $\phi$, one can rewrite this as \eqn\potagain{V(\phi)
= 2T_{3}\left(1 - { 1 \over 2 \pi^3}
 { T_{3}^3 \over  M_{10,Pl}^8 \phi^4} \right)~.}
 It was suggested in
\dvalitye\ that for large fields (large $r$), one may obtain
inflation from this potential.

A basic (and well known \quevedo) problem with this scenario is
the following.  The standard inflationary slow-roll parameters
$\epsilon$ and $\eta$ are defined via \eqn\epsis{\epsilon \equiv
{M_{Pl}^2 \over 2} ({V^\prime \over V})^2} \eqn\etais{\eta \equiv
M_{Pl}^2 {V^{\prime\prime}\over V}~.} One generally wants
$\epsilon, \eta \ll 1$ to get slow-roll inflation with sufficient
e-foldings. Is this possible in the model \potagain?  The
four-dimensional Planck mass appearing in \etais\ is $M_{Pl}^2 =
M_{10,Pl}^8 L^6$ where $L^6$ is the volume of M.  This implies
that $\eta$ is \eqn\etabec{ \eta = - { 10 \over \pi^3} (L/r)^6
\sim - 0.3 (L/r)^6 } Hence, $\eta \ll 1$ is possible only for $r >
L$ -- but two branes cannot be separated by a distance greater
than $L$ in a manifold $M$ of size $L$!

One can try to evade this constraint by considering anisotropic
extra dimensions or non-generic initial conditions which yield
flatter potentials than \potagain . We argue in Appendix A that
this is not possible. There are always some tachyonic directions
in the potential with $\eta \leq -2/3$. This implies that the
slow-roll approximation cannot be maintained for a large number of
e-foldings.

In \S3 we will explore another possibility that successfully
evades this problem -- we will modify the potential \potagain\ by
considering branes and antibranes in a warped geometry. We should
mention that there are other proposals which might evade the above
problem, such as branes at angles or branes with fluxes, see
\refs{\hhk,\Garcia,\shafi,\Kallosh,\tyeangle}.

However, all of these models have an unsolved problem: moduli
stabilization. For an internal manifold of  size $L$, the correct
four-dimensional Einstein-frame potential is not quite \potagain.
If one assumes that the main contribution to the inflationary
energy comes from the D3-brane tension then one finds, for $r
 \gg l_{s}$, that
\eqn\Vtrue{V(\phi,L) \sim {2T_{3} \over L^{12}}} The energy in the
brane tensions sources a steep potential for the radial modulus
$L$ of the internal manifold.
 Therefore, in the absence of a stabilization
mechanism which fixes $L$ with sufficient mass so that the
variation of $L$ in \Vtrue\ is negligible, one will find fast-roll
in the direction of large $L$ rather than slow-roll in the
direction of decreasing $r$. This means that it is important to
study concrete scenarios where the volume modulus has already been
stabilized.  However, we will show that not every means of volume
stabilization is compatible with inflation, even when the naive
inter-brane potential is flat enough to inflate.  We will return to the
issue of volume stabilization in \S5 , where we will discuss a new and
generic problem which appears when one considers the issue in
detail.

\newsec{Inflation in a Warped Background: Essential Features}

Our modified brane-antibrane proposal is that inflation might
arise from the interaction potential between a D3-brane and an
anti-D3-brane which are parallel and widely separated in
five-dimensional anti de Sitter space ($ \rm AdS_5$).\foot{This is
a slight simplification; in \S4 we will construct compact models
which deviate from $ \rm AdS_5$ both in the infrared and in the
ultraviolet.  It is nevertheless convenient to work out the
essential features of the model in this simpler geometry.}

The anti-D3-brane is held fixed at one location in the infrared
end of the geometry (this is naturally enforced by the dynamics,
as we shall explain). The D3-brane is mobile; it experiences a
small attractive force towards the anti-D3-brane. The distance
between the branes plays the role of the inflaton field.

The forces on the brane and antibrane arise as follows.  A single
D3-brane experiences no force in an AdS background: electrostatic
repulsion from the five-form background exactly cancels
gravitational attraction. The addition of a distant anti-D3-brane
results in a relatively weak interaction potential arising from
the attraction between the brane and the antibrane.  We interpret
this as a slowly varying potential for the inflaton.  We will
demonstrate in \S3.2 and in Appendix B that this potential is much
flatter than the interaction potential for a brane-antibrane pair
in flat space.

In the remainder of this section we explain this key idea in more
detail.  \S3.1 is a review of gravity in a warped background.
 \S3.2 deals with the motion of a brane probe in such a background.

It is important to point out that throughout this discussion, we
will ignore the possibility that other moduli (or the effects
which stabilize them) interfere with inflation.  In the context
of the string constructions of \S4, the relevant other modulus is
the compactification volume, and the generic problems associated
with its stabilization are the subject of \S5. In fact we will see
that this modulus problem will generically stop inflation.

\subsec{Gravity in an AdS Background}

We first consider a compactification of string theory on $\rm
AdS_5 \times X_{5}$ where $\rm X_{5}$ is a five-dimensional
Einstein manifold.\foot{The detailed form of $\rm X_{5}$ will not
matter for the moment.  For concreteness the reader may imagine
that $\rm X_5 = S^5$.} This arises in string theory as a solution
of ten-dimensional supergravity coupled to the five-form field
strength $F_5$.  The $\rm AdS_5$ solution is given in Poincar\'{e}
coordinates by the metric \eqn\metads{ds^2 = {r^2 \over
R^2}\left(-dt^2+{d\vec{x}}^2\right) + {R^2 \over r^2} dr^2} There
is, in addition, a five-form flux: if the geometry \metads\ arises
as the near-horizon limit of a stack of $N$ D3-branes, then the
five-form charge (in units of the charge of a single D3-brane) is
$N$.  $R$, the characteristic length scale of the $\rm AdS_5$
geometry, is related to the five-form charge by \eqn\ris{ R^4 =
4\pi a g_s N {\alpha^\prime}^2, } where the constant $a$ depends
on $X_5$. It will be useful to recall that $ \rm AdS$ is a
maximally symmetric, constant curvature spacetime. Its curvature
scales like $1\over{R^2}$ and is independent of the radial
location $r$.  As long as $N \gg 1$ this curvature is small and
supergravity analysis is reliable. We will choose to truncate $
\rm AdS_5$ to the region $r_0 < r < r_{max}$.

The reader will notice that, apart from the additional manifold
$\rm X_{5}$, this background is identical to that considered by
Randall and Sundrum in \RS. Two physical insights from \RS\ will
be crucial for our model.  First, one can see from the warped
metric \metads\ that the region of small $r$ is the bottom of a
gravitational well. Energies along the $t,x^i$ coordinates
therefore get increasingly redshifted as $r$ decreases.  (The
region of significant redshift is consequently referred to as the
infrared end of the geometry.)   Second, as a result of truncating
the AdS region, the four-dimensional effective theory which
governs low-energy dynamics will have a finite gravitational
constant, and will include four-dimensional gravity described by
the Einstein-Hilbert action: \foot{The graviton zero modes have
polarizations parallel to $t, x_i$, are constant on $\rm X_{5}$,
and have a profile identical to the warped background.}
\eqn\EH{S_{grav}= {1 \over 16 \pi G_N} \int d^4x \sqrt{-g}
\cal{R}.}

Recall also that in \RS, the truncation of AdS space was achieved
in a brute force manner by placing two branes, conventionally
called the Planck brane and the Standard Model brane, at $r_{max}$
and $r_0$, respectively. In the string theory constructions of
\GKP, the truncation arises because the compactification geometry
departs significantly from that of $ \rm AdS_5 \times X_{5}$ away
from the region $r_0 < r <r_{max}$.  In the ultraviolet, in the
vicinity of $r\geq  r_{max}$, the AdS geometry smoothly glues into
a warped Calabi-Yau compactification.  In the infrared, near $r =
r_0$, the AdS region often terminates
smoothly (as in the example of \KS).
 The infrared smoothing prevents the redshift factor
$r/R$  from decreasing beyond a certain minimum whose value will
be very important for our model.

\subsec{Brane Dynamics}

We mentioned above that  the warped nature of the geometry gives
rise to a redshift dependent on the radial location. It will be
important in the discussion below that the redshift results in a
very significant suppression of energies at the location of the
antibrane; that is, the ratio $r_0/R $ is very small.  Also note
that within the truncated AdS geometry, $r_0<r<r_{max}$, we have
chosen to place the anti-D3-brane at the infrared cutoff $r=r_0$,
where it has minimum energy due to the redshift effect.

The five-form background is given by
\eqn\valff{(F_5)_{rtx^1x^2x^3}={4 r^3 \over R^4}.} In a suitable
gauge the corresponding four-form gauge potential $C_4$ takes the
form \eqn\gaugepot{(C_4)_{tx^1x^2x^3}={r^4\over R^4}.}

The D3-brane stretches along the  directions $t, x^1,x^2,x^3$. Its
location in the radial direction of AdS space will be denoted by
$r_1$.  In the discussion below we will assume (self-consistently)
that the D3-brane has a fixed location along the angular
coordinates of the $\rm X_{5}$ space.  The motion of the D3-brane
is then described by the Born-Infeld plus Chern-Simons action
\eqn\dtact{S=-T_3\int \sqrt{-g} d^4x  \left({r_1^4 \over
R^4}\right) \sqrt{1-{R^4 \over r_1^4}g^{\mu\nu}
 \partial_\mu r_1 \partial_\nu r_1} +
T_3 \int (C_4)_{tx^1x^2x^3} dt dx^1dx^2dx^3.}  The indices
$\mu,\nu$ denote directions parallel to the D3-brane along the
$t,x^1,x^2,x^3$ coordinates, and $g^{\mu\nu}$ is the metric along
these directions.  The D3-brane tension, $T_3$, is
\eqn\valten{T_3={ 1\over (2 \pi)^3 g_s {\alpha^\prime}^2}.} For
future purposes we note here that since an anti-D3-brane has the
same tension as a D3-brane but opposite five-form charge, it is
described by a similar action where the sign of the second term is
reversed.

Now consider a  D3-brane slowly moving in the background given by
\metads\ and \valff, with no antibranes present.  It is easy to
see that because of a cancellation between the Born-Infeld and
Chern-Simons terms, the D3-brane action at low energies
 is just that of a free
field, \eqn\ffieldact{S= T_3 \int d^4x \sqrt{-g} {1 \over 2}
g^{\mu\nu}\partial_\mu r_1
\partial_\nu r_1.}  This in accord with our comment above that the
net force for a D3-brane in the background \metads,\valff\
vanishes due to gravitational and five-form cancellations.

We are now ready to  consider the effect of an antibrane on the
D3-brane.  Physically this arises as follows. The anti-D3-brane
has a tension and a five-form charge and  perturbs both the metric
and the five-form field. This in turn results in a potential
energy dependent on the location of the D3-brane.

The potential between a brane located at $r_1$ and an antibrane
located at $r_0$, in the limit when $r_1 \gg r_0$, is given  by:
\eqn\valpot{V=2T_3 {r_0^4 \over R^4}\left( {1- {1 \over N}{r_0^4
\over r_1^4}}\right).} For a derivation see Appendix B.

The first term in the potential is independent of the location of
the D3-brane and can be thought of as a constant potential energy
associated with the anti-D3-brane. It is proportional to the
tension $T_3$.  For the antibrane the force exerted by gravity and
the five-form field are of the same sign and add, so we have a
factor of 2.  In addition, the warped geometry gives rise to a
redshift, which reduces the effective tension of the antibrane by
a factor ${r_0^4 / R^4}$.

The second term in \valpot\ depends on the location of the
D3-brane; its negative sign indicates mutual attraction between
the pair. Two features of this term will be important in the
subsequent discussion.  First, the term varies slowly, as the
inverse fourth power of the radial location of the D3-brane.
Second, due to the warping of the background, the coefficient of
this second term is highly suppressed, by a redshift factor
$r_0^8/R^4$.

Two more comments are in order at this stage. We have assumed that
the antibrane is fixed  at $r_0$.  From \valpot, we see that this
is in fact a good approximation to make. In the $r_1 \gg r_0$
limit the first term in \valpot\ is much bigger than the second,
and most of the energy of the anti-D3-brane arises due to
interaction with the background. This is minimized when the
anti-D3-brane is located at $r_0$ in the truncated AdS spacetime.
Second, in our analysis above, we are working in the approximation
$r_1 \gg r_0$.  We will see below that the  D3-brane is far away
from the anti-D3-brane while the approximately sixty e-foldings of
inflation occur, so this condition is met during the inflationary
epoch. Eventually the D3-brane approaches the antibrane, $r_1 \sim
r_0$, and this approximation breaks down.  The potential then
becomes quite complicated and more model dependent (e.g. it
depends on the separation between the brane and antibrane along
$\rm X_{5}$ ).  The resulting dynamics is important for reheating.

A  summary of the discussion so far is as follows.  We have
considered a D3-brane moving in an $ \rm AdS_5\times X_5$
background with five-form flux, in the presence of a fixed
anti-D3-brane. This system is described by an action:
\eqn\fullact{S= \int d^4x
\left( {1 \over 2}T_3 g^{\mu\nu}\partial_\mu r_1
\partial_\nu r_1 -  2 T_3{r_0^4 \over R^4} \left({1- {1 \over N}
{r_0^4\over r_1^4}}\right) \right)  }
 The reader will notice in particular
that $r_1$, the location of the D3-brane, is a scalar field in the
effective four-dimensional theory.

Once we cut off the $\rm AdS_5$ space as in the Randall-Sundrum
models we will find that we can add to \fullact\ the
four-dimensional Einstein action. However, we should also add an
extra coupling of the form $ {T_3 \over 12}r_1^2 {\cal R}$ coming
from the fact that the scalar field $r_1$ describing the position
of the D3-brane is a conformally coupled scalar \SeibergWitten .
This unfortunately leads to a large contribution to $\eta$.  We
will discuss this phenomenon in more generality (from the
perspective of the effective low-energy four-dimensional
supergravity) in \S5.

The model described above has several appealing features in
addition to the flatness of the potential.  We study these
properties in Appendices C,D, and E, with the assumption that one
can somehow overcome the problems of \S5\ (which must be
tantamount to cancelling the conformal coupling).  In Appendix C
we compute the inflationary parameters and show that observational
constraints are easily met. In Appendix D we argue that eternal
inflation can be embedded into this model, and in Appendix E we
point out that the warped geometry suppresses the production of
metric perturbations due to cosmic strings (which naturally form during
the brane/anti-brane annihilation).

\newsec{A Concrete Example in String Theory}

We now show how to realize our proposal in a specific class of
string compactifications.  In \S4.1 we present the
compactifications and explain why they contain warped throat
regions.  As the warped throat is well-described by the
Klebanov-Strassler (KS) solution \KS, we dedicate \S4.2 to a very
brief review of the KS geometry. In \S4.3 we show that a brane
moving in the KS background might give
 rise to inflation, realizing the
general idea presented in \S3. Throughout this discussion, we
ignore the problem of stabilizing the overall volume modulus,
which is unfixed in the constructions of \GKP.  We consider the
problem of volume stabilization in \S5, where we will find that
generic methods of volume stabilization can perturb the inflaton
enough to stop inflation.

\subsec{The Compactification}

Our starting point is type IIB string theory compactified on a
six-dimensional Calabi-Yau orientifold.  More generally one could
use F-theory on an elliptically-fibered Calabi-Yau fourfold. We
choose to turn on background fluxes: the three-form fluxes $F_3,
H_3$ present in the theory are placed along cycles in the internal
space (and $F_5$ is fixed as in \GKP).  These fluxes induce
warping of the background.  One can show that the resulting space
is a warped product of Minkowski space and the Calabi-Yau:
\eqn\warp{ ds^{2} =
e^{2A(y)}\eta_{\mu\nu}dx^{\mu}dx^{\nu}+e^{-2A(y)}g_{mn}dy^{m}dy^{n}}
where $y_i$ are coordinates on the compactification manifold and
$g_{mn}$ is the Calabi-Yau metric. As was discussed in \GKP, one
expects that with a generic choice of flux, all the complex
structure moduli of the Calabi-Yau, as well as the dilaton-axion,
will be fixed. We will assume that the compactification has only
one K\"ahler modulus, the overall volume of the internal space.

As described in \GKP, one can use the above construction to
compactify the warped deformed conifold solution of Klebanov and
Strassler (KS).  We spend the next section reviewing a few facts
about this geometry, as certain details will be important for
inflation.

\subsec{The Klebanov-Strassler Geometry}

The Klebanov-Strassler geometry \KS\ is a noncompact
ten-dimensional solution to type IIB supergravity in the presence
of background fluxes.  The spacetime naturally decomposes into a
warped product of a Minkowski factor and a six-dimensional
internal space.  The six-dimensional space has a tip which is
smoothed into an $S^3$ of finite size. Far from this tip the
geometry can be approximated by a cone over the Einstein manifold
$T^{1,1}$, which is topologically $S^2\times S^3$.  Our
coordinates will be five angles on $T^{1,1}$, which we can
consistently neglect in the following, and a radial coordinate $r$
which measures distance from the tip.  The background fluxes are
given by \eqn\flux{{1\over {(2\pi)^2 \alpha^\prime}}\int_{A} F =
M, ~~~~~~~~~ {1\over {(2\pi)^2 \alpha^\prime}}\int_{B} H = -K}
where A is the $S^3$ at the tip and B is its Poincar\'{e}-dual
three-cycle. We will require that $M \gg 1$ and $K \gg 1$; these
conditions are important in deriving the solution. The exact
metric is known, but for our purposes a simpler form, valid far
from the tip, will be more useful.  For large $r$ we may express
the complete ten-dimensional solution as \eqn\warpks{ ds^2 =
h^{-1/2} \eta_{\mu\nu}dx^{\mu} dx^{\nu} + h^{{1/2}}{\left({dr^2 +
{r^2}ds^2_{T^{1,1}} }\right)}} where now \eqn\limh{ h(r) ={{27
{\pi}}\over{4r^4}}{\alpha^\prime}^2 g_s M\left(K + {g_{s}M}
\left({3\over{8 \pi}}+ {3\over{2\pi}}\rm{
ln({r\over{r_{max}}})}\right)\right).} Neglecting the logarithmic
corrections and the second term  on the right, this takes the form
\foot{The second term on the r.h.s. of \limh\ can easily be
included. For the numerical values discussed in Appendix C, this
gives a three percent correction.} \eqn\defr{R^4 = {27 \over 4}\pi
g_s N {\alpha^\prime}^2} \eqn\defn{N \equiv M K} When the KS
geometry is embedded in a compactification then at some location
$r = r_{max}$ the warped throat geometry is smoothly joined to the
remainder of the warped Calabi-Yau orientifold. Near this gluing
region, departures from the $ \rm AdS_5 \times T^{1,1}$ geometry
are noticeable; eventually the $ \rm AdS$ must end.  In terms of
redshift this location corresponds to the deep ultraviolet, and so
the gluing region plays the role of the ultraviolet cutoff (Planck
brane) in the $ \rm AdS$ of \S3.

The exact solution likewise shows departures from \warpks\ in the
far infrared, near the tip, although the geometry remains
smooth.\foot{The radius of curvature is $\sqrt{g_s M
\alpha^\prime}$, so the tip is smooth provided $g_s M \gg 1$.} The
details of the deviation from \warpks, although known, are
unimportant here; it will suffice to know the redshift at the tip.
This can be modeled by cutting off the radial coordinate at some
minimum value $r_0$, which is the location of the tip.  It was
shown in \GKP\ that the minimal redshift satisfies
\eqn\maxred{{r_0\over{R}} = e^{-{2 \pi K \over 3 g_s M}}} This can
be extremely small given a suitable choice of fluxes.

 \subsec{Inflation from Motion in the KS Region}

In \KKLT\ additional anti-D3-branes were introduced in the KS
region. These anti-D3-branes minimize their energy by sitting at
the location where the redshift suppression is maximum, i.e. at
the very tip of the deformed conifold, where $r\sim r_0$ (the
dynamics of anti-D3-branes in the KS geometry was studied in
\KPV).

Thus we see that the string construction outlined above has all
the features of the general model of \S3: a truncated $ \rm AdS_5$
geometry, an associated five-form flux of the correct strength,
and anti-D3-branes fixed at the location of maximum redshift.  In
addition most of the moduli associated with the compactification,
including the dilaton, are stabilized. The one exception is the
volume modulus; we will discuss the complications its
stabilization introduces separately, in \S5.

No mobile D3-branes were included  in the construction of \KKLT,
but it is easy to incorporate them.  One needs to turn on somewhat
different values of three-form flux, which allow the four-form
tadpole to cancel in the presence of the additional D3-branes.
This is straightforward to do and does not change any of the
features discussed above.

We will take one such D3-brane to be present in the KS region of
the compactification.  The general discussion of \S3 applies to
this brane. Since the D3-brane is described by the action \dtact,
with $R$ now given by \defr, the calculation of the
brane-antibrane potential follows the discussion in Appendix C,
which we outline here.

 In the KS model the warp factor  \warp\ is
given in terms of a function $h \equiv e^{-4 A}$ which obeys a
Laplace equation, with the fluxes and branes acting as sources. In
particular, a single D3-brane located at $r=r_1$ will correct the
background according to \eqn\newh{ h_{new}(r) = h(r) + \Delta
h(r,r_1).}  Here $h(r)$ is the background given in \limh\ and
$\Delta h(r,r_1)$ is the correction due to the D3-brane. In a
region where the original warp factor is very small we see that
$h(r_0) \gg 1$, so that the total warp factor can be expanded as
\eqn\newwarp {e^{4 A} \sim h(r_0)^{-1} \left( 1 - {\Delta h(r,r_1)
\over{ h(r_0)}}\right).} This warp factor yields the contribution
to the energy due to the presence of an antibrane. If $h(r_0) \gg
1$ this typically gives a very flat potential.

The small warp factor and the consequent exponential flatness are
the heart of our proposal, so an alternative explanation of the
origin of these small numbers may be helpful.  Recall that there
is a holographic dual gauge theory which describes the geometry of
the KS model. This gauge theory is approximately scale invariant
in the deep ultraviolet, with slowly running gauge couplings. It
undergoes $K$ duality cascades before leading in the infrared to a
confining gauge theory with a mass gap. Then the smallness of the
redshift factor,\eqn\maxredtwo{{{\left({r_0\over{R}}\right)}^4} =
e^{-{8 \pi K \over 3 g_s M}}} can be ascribed to the exponential
smallness of the confinement scale in such a gauge theory.

In summary, we have seen that one can construct concrete examples
of string compactifications which lead to the general behavior
described in \S3.  One of their virtues is that they automatically
lead to very flat inflaton potentials, without the need for large
brane separation or excessive fine-tuning of initial conditions.
The primary source of this flatness is the redshift suppression
\maxred\ which is exponentially sensitive to the (integer) choice
of fluxes $K$ and $M$.  However, all of these virtues must be
re-examined in the light of concrete ideas about how to stabilize
the closed string moduli.  In this general class of flux
compactifications, the fluxes stabilize many moduli but not e.g.
the overall volume. We now turn to the discussion of volume
stabilization.

\newsec{Volume Stabilization: New Difficulties for D-brane
Inflation}

The results of \S3,4\ indicate that warped geometries provide a
promising setting for making models of inflation with naturally
small $\epsilon$ and $\eta$.  However, as emphasized in \S2, one
must ensure that the compactification volume is stabilized in
order to avoid rapid decompactification instead of inflation. We
will now demonstrate that in the concrete models of \GKP\ this is
far from a trivial constraint.

In these models the four-dimensional ${\cal N}=1$ supergravity at
low energies is of the no-scale type.  The K\"ahler potential for
the volume modulus $\rho$ and the D-brane fields $\phi$ takes the
form \dWG\foot{The variable $\rho$ is called $-i\rho$ in \KKLT.}
\eqn\nosc{K(\rho,\bar\rho,\phi,\bar\phi) = - 3 \log \left(\rho +
\bar \rho  - k(\phi,\bar\phi)\right)~} Let us pause for a moment
to explain how this is obtained. In the tree level
compactification the massless fields are the volume, the axion and
the position $\phi$ of the branes. The axion comes from a
four-form potential proportional to a harmonic four-form in the
internal manifold \GKP. At first sight one would think that the
moduli space is simply a product of the moduli space for $\phi$,
which is just the internal Calabi-Yau manifold, and the space
spanned by the volume and the axion. This is not correct; the
axion describes a circle which is non-trivially fibered over the
$\phi$ moduli space. This structure arises from the coupling of
the four-form potential to the worldvolume of the moving D3-brane.
The moduli space has a metric of the form \eqn\metricreal{ ds^2 =
{3\over{2r^2}}\left(  dr^2 + (d \chi + {1 \over 2} i k_{, j} d
\phi^j - {1 \over 2} i k_{, \bar j} d \bar \phi^j )^2 \right)  + {
3 \over  r} k_{, i \bar j} d \phi^i d \phi^{\bar j} } where $r$ is
proportional to the volume of the Calabi Yau (in the notation of
\GKP, $r \sim e^{4u}$).
 If we tried to work with a complex variable $ r + i \chi  $ then
\metricreal\ would not follow from a K\"ahler potential. It turns
out that the good complex variable is $\rho$, which is defined as
follows. The imaginary  part of $\rho$ is the axion, while the
real part of $\rho$ is defined by \eqn\volis{ 2r = \rho + \bar
\rho  - k(\phi,\bar\phi)~.} It is then possible to see that \nosc\
gives rise to \metricreal . This type of definition of $\rho$
arises when we Kaluza-Klein compactify supergravity theories; see
for example \wittenred .

 The superpotential is of
the form \eqn\wis{W = W_0} where $W_0$ is a constant (we assume
the D-branes are on their moduli space, so we do not write down the
standard commutator term). This arises from the (0,3) part of the
three-form flux in the full theory including the complex structure
moduli and the dilaton. We have not yet included the
anti-D3-branes used in \S3,4; these will be incorporated at the
end of the discussion.

It is important that with the K\"ahler potential \nosc, one
obtains the no-scale cancellation in the potential
 \eqn\cancelt{ V = e^{K}( g^{a\bar b}K_{,a}K_{,\bar b}
|W|^2 - 3|W|^2) = 0~.}

since
\eqn\cancel{g^{a\bar b}\partial_{a}K \partial_{\bar b}K
=
3
} where $a,b$ run over $\rho$ and $\phi$. \foot{ The easiest way
to check \cancel\
 is to note that in expression \cancel\ we can switch back to
the variables $r,a,\phi$ in \metricreal. In these variables $K$ is
only a function of $r$. }

Using \cancelt, it is clear that a generic $W(\phi)$ will yield a
potential for the D-brane fields $\phi$, but that the potential
for the $\rho$ modulus   will vanish if the solution for the
$\phi$ fields has $\partial_{\phi}W = 0$. It is also clear that a
${\it constant}$ superpotential, as in \wis, gives no potential to
the $\phi$ fields.  This is consistent with the analysis in \GKP,
where the pseudo-BPS nature of the flux background leaves the
D3-brane moduli unfixed.

We are interested in finding a situation where the
D-branes can move freely in the Calabi-Yau (so the $\phi$ fields
are ${\it unfixed}$), but the volume is stabilized.  Before we
discuss various scenarios for such a stabilization, it is
important to distinguish carefully between the $\rho$ chiral
superfield, and the actual volume modulus, $r$,  which controls the
$\alpha^\prime$ expansion.

The K\"ahler potential \nosc\ has the following peculiar feature.
Let us imagine that there is one D-brane, and hence a triplet of
fields $\phi$ describing its position on the Calabi-Yau space.
Then $k(\phi,\bar\phi)$ should be the K\"ahler potential for the
Calabi-Yau metric itself, at least at large volume.  However,
under K\"ahler transformations of $k$, the expression \nosc\ is
not well behaved.  This can be fixed by assigning the
transformation laws \eqn\kahtrans{k(\phi,\bar\phi) \to k + f(\phi)
+ \overline{f(\phi)},~~ \rho \to \rho + f, ~~\bar\rho \to \bar\rho
+  \bar f~.} This is a manifestation of the fact that the circle
described by the axion is non-trivially fibered over the $\phi$
moduli space. Note that the physical volume of the internal
dimensions, which is given by  $r$,  \volis , is invariant under
\kahtrans.

Armed with this knowledge, and given \nosc\ and \wis\ as our
starting point, we can now explore various scenarios for volume
stabilization.

\subsec{Scenario I: Superpotential Stabilization}

Perhaps the most straightforward method of stabilizing the volume
involves a nonperturbative contribution to the superpotential.
Various sources of nonperturbative superpotentials for the $\rho$
modulus are known; one instructive example described in \KKLT\
involves a superpotential \eqn\wnpis{ W(\rho) =W_{0} + A e^{-a
\rho} } where $A$ and $a$ are constants and $W_{0}$ is the
contribution \wis\ of the three-form flux.  For the remainder of
this section we will consider $W=W(\rho)$ to be a general
holomorphic function of $\rho$.

In the presence of D3-branes the superpotential must in addition
develop some dependence on $\phi$, as it should be invariant under
\kahtrans . For instance, as argued in \ganor, the superpotential
due to Euclidean brane instantons or gauge dynamics on D7-branes
has to vanish when a D3-brane hits the relevant cycle. This can be
understood directly by examining and integrating out the massive
D3-D7 strings in the latter case. This subtlety must be accounted
for to get a globally well-defined $W$, and we will see in a
moment that this actually changes the inflaton mass term.
Nevertheless, we will first study the simpler case $W=W(\rho)$,
both because it reflects the essential features of the problem and
because the full dependence of $W$ on $\phi$ is not known.

Let us start by presenting a general argument which highlights a
problem faced by  any inflationary model involving a moving
D3-brane in the models of \KKLT . The main point is that one will
choose some configuration with a positive energy $V$. When the
compact manifold is large then this energy will go to zero rather
quickly, as a power of the volume modulus $r$: \eqn\vandr{
V(r,\phi) = {X(\rho)\over{r^{\alpha}}} = {X(\rho)\over{(\rho -
\phi \bar \phi /2)^{\alpha}}}} where $\alpha$ is a number of order
one and the form of $X(\rho)$ depends on the source of energy.
This follows because in existing proposals the inflationary energy
arises either from brane tensions or from fluxes, and all known
brane and flux energies vanish as some power of $r$.
On the other hand the stabilization mechanism would fix $\rho$ (or
else some combination of $\rho$ and $\phi)$ rather than $r$. This
implies that as the brane moves and $\phi$ changes there will be a
change in the potential, \eqn\newdeltavis{ V = V_0\left(1+\alpha
{\phi \bar \phi \over 2 r}+ ...\right).} This will lead to a
contribution to $\eta$ of order one, unless there is a
compensating contribution to the mass term from some other source.

One possible source of such a cancellation is a dependence of the
superpotential on $\phi$, not just $\rho$.  If $V(r,\phi) =
{X(\rho,\phi)r^{-\alpha}}$ then we would get an additional
contribution to the mass term,
$$V(\rho, \phi) = {X(\rho, \phi)\over (\rho -  \phi \bar \phi /2)^{\alpha}}=
 {X(\rho )\over \rho   ^{\alpha}}\left(1+\alpha {\phi \bar \phi \over 2 r}+...\right)+
 {\Delta(\rho)\over r^{\alpha}}\phi \bar \phi $$
where
$$X(\rho, 0)\equiv X(\rho) \qquad \Delta(\rho)\equiv
\partial_{\phi} \partial _{\bar \phi} X(\rho, \phi)|_{\phi=0}$$
so that at the minimum $\rho = \rho_c$ we find
$$V(\rho_{c}, \phi) =  V_0(\rho_{c}) +\left ( {\alpha V_0(\rho_{c})\over 2 \rho_{c}}
 + {\Delta(\rho_{c})\over \rho_{c}^{\alpha}}\right)\phi \bar \phi +...$$
In principle the second contribution to the mass term might
substantially cancel the first, alleviating the problem of the
inflaton mass.  This would certainly require fine-tuning at the
level of one percent (in order to make $\eta$ sufficiently small
to allow sixty e-foldings).  More importantly, the dependence of
$W$ on $\phi$ is not known, so the question of which models admit
such fine-tuning cannot be answered at present. We should
emphasize that the problem we are discussing is quite general, but
one might well be able to find non-generic configurations in which the
problem is absent.

Let us discuss these issues more concretely for the case of a
brane-antibrane pair transverse to a stabilized Calabi-Yau.  In
principle one should be able to compute the inflaton potential
directly, by substituting the complete superpotential into the
supergravity F-term potential \eqn\potnow{V^{F} = e^K (g^{i\bar j}
D_i W \overline{D_j W} - 3 |W|^2) } and possibly including the
effects of D-term contributions.  This turns out to be a rather
subtle problem, essentially because of the breaking of
supersymmetry in the brane-antibrane system.

We will begin instead by understanding the (supersymmetric) system
of a single D3-brane transverse to a Calabi-Yau.  We will find
that superpotential stabilization of the volume necessarily
generates mass terms for the scalars $\phi$ which describe the
motion of the D3-brane.  An implicit assumption in brane-antibrane
inflation scenarios is that the brane and antibrane are free, in
the absence of interactions, to move around the Calabi-Yau; the
gentle force from their Coulomb interaction is then expected to
lead to a relatively flat inflaton potential. Significant mass
terms for the D3-brane (or any external forces on the D3-brane)
invalidate this assumption and make inflation impossible.

Let us therefore consider the effective potential governing a
D3-brane transverse to a Calabi-Yau manifold.  We substitute the
superpotential $W(\rho)$ and the K\"ahler potential \nosc\ into
\potnow, where the physical volume modulus $r$ is given by \volis.
The resulting four-dimensional effective potential is
\eqn\visnow{V^{F} = {1\over 6 r}\left( \partial_{\rho}W
\overline{\partial_{\rho}W} (1 + {1\over {2
r}}{k,_{\phi}k,_{\bar\phi} \over {k,_{\phi\bar\phi}}}) - {3\over
{2 r}}(\overline{W}\partial_{\rho}W + W
\overline{\partial_{\rho}W})\right)~.} In the vicinity of a point
in moduli space where $k(\phi,{\bar \phi})=\phi {\bar \phi}$, this
can be simplified to \eqn\vsimpler{V^{F} = {1\over 6 r}\left(
|\partial_\rho W|^2  - {3\over {2 r}}(\overline{W}\partial_{\rho}W
+ W \overline{\partial_{\rho}W})\right) +  \left({|\partial_\rho
W|^2\over{ 12 r^2}}\right)\phi {\bar \phi} ~.}

We must now incorporate the effects of an anti-D3-brane.  In the
scenario of \KKLT\ the superpotential \wnpis\ stabilized the
compactification volume and generated a negative cosmological term
$V_0$.  The positive, warped tension of an anti-D3-brane was added
to this to produce a small positive cosmological constant.  In our
notation, the anti-D3-brane induces an additional term in the
effective potential \visnow, \eqn\nowvis{V = {1\over 6 r}\left(
\partial_{\rho}W \overline{\partial_{\rho}W} (1 + {1\over {2
r}}{k_{,\phi}k,_{,\bar\phi} \over {k_{,\phi\bar\phi}}}) - {3\over
{2 r}}(\overline{W}\partial_{\rho}W + W
\overline{\partial_{\rho}W})\right) + {D\over{(2r)^2}}} where $D$
is a positive constant.  Notice that this induced term differs
from the one in \KKLT\ by a factor of $r$.  This arises because
the anti-D3 tension in the warped compactifications of \GKP\
scales like ${1\over r^3} e^{4A}$, and in the highly warped
regime, $e^{4A} \sim r ~exp(-{{8\pi K}\over {3g_s M}})$.  This
does not alter the conclusions of \KKLT, though it changes the
numerology.

Suppose that the potential \nowvis\ has a de Sitter minimum
$V_{dS}$ at $\rho=\rho_c, \phi,{\bar \phi}=0$.  We will now
compute the mass of the D3-brane moduli in an expansion about this
minimum. To simplify the analysis we assume that at the minimum
$\rho$ is real, and also that for real $\rho$, $W(\rho)$ is real.
The canonically normalized scalar which governs the motion of the
D3-brane is not $\phi$ but is instead a rescaled field $\varphi =
\phi \sqrt{{3/{(\rho+\bar\rho)}}}$; it is the mass of $\varphi$
which we will compute.

First, we rewrite \nowvis\ as \eqn\nextv{ V =
\left({{W'(\rho)^2\rho - 3W(\rho)W'(\rho)+{D\over{4}}
}}\right){(\rho-\phi\bar\phi/2)^{-2}}} where primes denote
derivatives with respect to $\rho$, and define $V_0$ by
\eqn\vzerois{ V_0(\rho_c) =
{1\over{\rho_c^2}}{\left({{W'(\rho_c)^2\rho_c -
3W(\rho_c)W'(\rho_c)+{D\over{4}}}}\right)}.} Then \eqn\vexpand{ V
= {V_0(\rho_c)\over{(1-\varphi\bar\varphi/3)^2}}\approx
V_0(\rho_c)\left( 1+ {2\over 3} \varphi \bar \varphi\right).} This
means that the  field $\varphi$  acquires the mass \eqn\conda{
m^2_\varphi ={2\over 3}V_{dS} = 2H^2}

This is in fact precisely the result one would obtain for a
conformally coupled scalar in a spacetime with cosmological
constant $V_{dS}$.  This is most easily understood by setting
$D=0$ and studying the resulting $\rm AdS_{4}$.  The
four-dimensional AdS curvature is $R_{AdS} = 4 V_0$, so that
\conda\ corresponds to a coupling \eqn\coupling{{\delta V =
\left({1\over{6}}R_{AdS}\right) \varphi \bar \varphi}.} If the
D3-brane is in a highly warped region this result could have been
anticipated, since this highly warped region is dual to an almost
conformal four-dimensional field theory \juanAdS\ and the scalar
field describing the motion of the brane is conformally coupled
(see \SeibergWitten )\foot{Note that the kinetic term for
$\varphi$ is of the form $\int d^4x  \nabla \varphi \nabla \bar
\varphi $.}. The derivation of \coupling\ is also valid even when
the D3-brane is far from the near horizon region.

We now see that the D3-brane moduli masses are necessarily of the
same scale as the inflationary energy density $V_0$, since during
inflation, the extra antibrane(s) simply sit at the end of the
throat and provide an energy density well-modeled by \nowvis.  It
is straightforward to verify that such masses lead to a slow-roll
parameter  $\eta = 2/3$, incompatible with sustained slow-roll
inflation.

It is instructive to compare this result with the well-known
$\eta$-problem, which bedevils most models of F-term inflation in
${\cal{N}}=1$ supergravity.  One begins by asking whether
slow-roll inflation is possible in a model of a single field
$\phi$ with any type of K\"ahler potential and any superpotential
$W(\phi)$.  For a minimal K\"ahler potential and a generic
superpotential $W(\phi)$ one typically has a inflaton mass
$m_\phi^2 = {\cal{O}}(H^2)$, and hence no inflation, just as in
the generic case considered in the present paper. But this does
not mean that inflation in ${\cal{N}}=1$ supergravity is
impossible. Various superpotentials with non-generic dependence on
$\phi$ have been found, some of which permit inflation. For
example, in supergravity with the canonical K\"ahler potential and
a linear superpotential for the inflaton, the mass term
contribution to the potential cancels: \eqn\Fterm {K=\bar \phi
\phi\ ,  \qquad W= \phi \  \qquad \Rightarrow \quad V= e^{\bar
\phi \phi}\left((1+\bar \phi \phi )^2-3 \bar \phi
\phi\right)=1+{1\over 2} (\bar \phi \phi)^2+\cdots} A similar
effect occurs for the superpotential $W =
\phi(\sigma_1\sigma_2-M^2)$, which leads to a simple realization
of F-term hybrid inflation \fterm. Moreover, the dangerous mass
terms for the inflaton do not appear at all in D-term inflation
\BinetruyXJ.

It is quite possible, therefore, that one could find a consistent
inflation scenario in string theory by studying superpotentials
which depend on the inflaton field.  As mentioned above, this
would undoubtedly require a fine-tuned configuration in which two
contributions to the mass cancel to high precision.  We treat this
question in detail in Appendix F, where we show that the
introduction of a superpotential depending on the inflaton field
$\phi$ leads to a modification of the mass-squared $m_{\phi}^2$ of
the inflaton field which could  make it much smaller (or much
greater) than $2H^2$. This issue merits further investigation,
which should become possible as we learn more about the detailed
dependence of $W(\rho,\phi)$ on the background geometry and on the
fluxes in string compactifications.

\subsec{Scenario II: K\"ahler Stabilization}

One model of stabilization that would be compatible with the
inflationary scenario of \S3,4 is the following.  We have seen
that the true K\"ahler-invariant expansion parameter which
controls the $\alpha^\prime$ expansion in these models, is $r$.
Furthermore, $r$ and $\phi$ have independent kinetic terms.

A method of directly stabilizing $r$ could freeze the
volume directly, without stopping inflation.  Since $r$ is
not a chiral superfield itself, stabilization via effects in the
superpotential cannot accomplish this.  However, given that $W_0
\neq 0$, one ${\it can}$ imagine that corrections to the K\"ahler
potential could directly stabilize $r$.

In fact, K\"ahler stabilization has been proposed earlier for
rather different reasons (see e.g. \banksdine, which discusses
K\"ahler stabilization of the heterotic string dilaton).  Here we
would need the $\alpha^\prime$ corrections to \nosc\ to break the
no-scale structure and fix $r$.  Some of these corrections have
been calculated (see e.g. \bbhl). The subset of terms presented in
\bbhl\ does not lead to this kind of stabilization, though there
are likely to be other terms at the same orders which could change
this conclusion.  However, K\"ahler stabilization would be very
difficult to find in a controlled calculation, so one might simply
have to state it as a model-building assumption.

If one does assume that $r$ is stabilized by corrections to the
K\"ahler potential, then the models of \S3,4 could be realized in
the framework of \GKP.  In Appendix C we show that in these models
one can easily satisfy observational constraints such as the
number of e-foldings and the size of the density perturbations.

\newsec{Conclusion}

One of the most promising ideas for obtaining inflation in string
theory is based on brane cosmology. However, brane-antibrane
inflation \dvalitye\ suffers from various difficulties when one
tries to embed it in full string compactifications with moduli
stabilization, such as the (metastable) de Sitter vacua of \KKLT.

We have argued here that some of these difficulties can be
resolved by introducing highly warped compactifications.  The
warped brane-antibrane models introduced in general form in \S3
and in a compact string theory example in \S4 give rise to
slow-roll inflation with an exponentially flat potential.  In the
compact example, the slow-roll parameters and the density
perturbations can be fixed at suitable values by an appropriate
choice of discrete fluxes in the warped region.

The above discussion assumes a suitable stabilization mechanism
for the volume modulus of the compactification manifold. As
described in \S5, this is a highly nontrivial issue. Indeed, we
have found that if one stabilizes the moduli as in \KKLT\ then
this field acquires an effective mass-squared $m_\phi^2 =
{\cal{O}}(H^2)$, making inflation impossible.  As discussed in
\S5.1, fine-tuned dependence of the superpotential on $\phi$ could
reduce this mass.  With {\it{generic}} dependence on $\phi$ the
problem persists.

The arguments leading to our conclusion that generic methods of
stabilization stop inflation are rather general, and should apply
to any system where the energy density depends on the volume
modulus as $r^{-\alpha}$ with $\alpha > 0$.  There are general
arguments that this should always be the case, for the sources of
energy we know about in string theory \Giddings.
Thus, it appears very
difficult to achieve slow-roll brane inflation in a manner
compatible with stabilization of the compactified space in string
theory.  At the very least, it is challenging to find a model
which works for {\it{generic}} forms of the stabilizing
superpotential, which itself varies in a way that depends on all
of the microscopic details of the compactification.  In those
non-generic cases where inflation is possible, the inflationary
predictions will depend on the details of the moduli
stabilization.

One should note that the degree of fine-tuning required for
slow-roll inflation in these models is not extraordinary (see
Appendix F), and may well be attainable within the large class of
known models. Moreover, even though fine-tuning is certainly
undesirable, it may not be a grave problem. Indeed, if there exist
many realizations of string theory, then one might argue that all
realizations not leading to inflation can be discarded, because
they do not describe a universe in which we could live.
Meanwhile, those non-generic realizations which
lead to eternal inflation (see Appendix D) describe inflationary
universes with an indefinitely large and ever-growing volume of
inflationary domains. This makes the issue of fine-tuning less
problematic.  It will not escape the reader's notice that this
argument is anthropic in nature \refs{\BP,\andreian,\lenny}. It is worth
pointing out that it is an independent, presumably well-defined
mathematical question, whether or not string theory has
solutions which are consistent with present experiments
(e.g. which contain the standard model of particle physics,
have sufficiently small cosmological term, and allow early
inflation).  This question can of course be studied
directly (see e.g. \Douglas\ for recent
work in this direction), and is an important one for string
theorists to answer.  Only if string theory does admit such
solutions, does anthropic reasoning in this context become tenable.
The large diversity of string vacua makes it reasonable
to be optimistic on this score.

We have primarily focused on the implications of superpotential
stabilization of the moduli for D3-brane/anti-D3-brane inflation.
Our analysis has implications for other models of brane inflation
as well.  These include $Dp-{\overline{Dp}}$ systems and
$Dp$-branes at angles with $p=5,7$. In these cases, Chern-Simons
couplings will generically induce a $D3$-brane charge on the
branes due to the presence of a non-trivial $B_{NS}$ field. Such a
charge will also be generated due to the curvature couplings for
generic topologies of the cycles the branes wrap. If the induced
charge is of order unity or more, the discussion of the previous
section will apply. The volume modulus and the inflaton field will
mix non-trivially in the K\"ahler potential and as a result a
superpotential of the kind considered in \S5.1, or in fact any
source of energy which scales like $1/r^\alpha$,  will generically
impart an unacceptably big mass to the inflaton. It would be
interesting to explore the special cases where such a charge is
not induced, to see if one can make simple working models of brane
inflation.

Other existing proposals for brane inflation depend on
Fayet-Iliopoulos terms in the low-energy field theory \BinetruyXJ .
The status
of these FI  terms in the effective ${\cal{N}}=1$ supergravity arising
from compactified string theory therefore merits careful
investigation.
String theory models with D-terms were
realized in brane constructions \refs{\hhk,\Kallosh} without
consideration of volume stabilization. A consistent embedding of
this model into compactified string theory is under investigation
\kalletal.

\bigskip
\centerline{\bf Acknowledgements}

We would like to thank C. Burgess, M. Dine, M. Fabinger,
A. Guth, L. Kofman, S. Prokushkin, F. Quevedo, N. Seiberg, A. Sen, M. M.
Sheikh-Jabbari, S. Shenker, E. Silverstein, L. Susskind, S.
Thomas, H. Tye and E. Witten for helpful discussions.  This
research was supported by NSF grant PHY-9870115.  The work of S.K.
was also supported in part by a David and Lucile Packard
Foundation Fellowship for Science and Engineering and the DOE
under contract DE-AC03-76SF00515. The research of JM is supported
in part by DOE grant DE-FG02-90ER40542.  The work of L.M. was
supported in part by a National Science Foundation Graduate
Research Fellowship. S.P.T. acknowledges
support from the Swarnajayanti Fellowship,
DST, Govt. of India, and most of all from the people of India.

\appendix{A}{General Discussion of Brane-Antibrane Potentials}

Here we compute the gravitational force between a D3-brane and an
anti-D3-brane which are transverse to a general compact
six-dimensional space.
We assume that there is no warping before
we add the D-branes. Our objective is to compute the expression
for the slow-roll parameter $\eta$ \etais\ in this setup. For this
purpose we note that the brane tension as well as the
ten-dimensional Planck mass drop out from the expression for
$\eta$ if we express it in terms of the physical distance. We can
therefore set $M_{Pl, 10} =1$, $T_{D3} =1$, to avoid clutter in
the equations.

The action for the system has the form \eqn\action{ S = \int d^6x
{ 1 \over 2} ( \nabla \varphi)^2 +   \sum_i
 (1 + \gamma \varphi(x_i))
} where $\gamma$ is a constant we will determine below. Here
$\varphi$ is the gravitational potential on the internal space.
The equation of motion from \action\ is \eqn\motac{ - \nabla^2
\varphi + \gamma \sum_i \delta(x-x_i) =0 } Treating one brane as
the source and the other as a probe and comparing with \bantibpot\
we see that $\gamma^2 = 2$.\foot{ Note that \bantibpot\ contains a
contribution both from gravity and from the Ramond fields, so the
gravity contribution is half of that in \bantibpot .} The
expression for the energy of $N$ branes is thus \eqn\expres{ V
\sim N + {1 \over 2} \sum_i  \gamma \varphi_{others}(x_i) } where
the subscript in $\varphi$ indicates that we compute the potential
due to the other branes, with $j\not =i$, and evaluate it at
$x_i$. There is also a self-energy correction. We assume that the
latter is independent of position. This is true in homogeneous
spaces, such as tori.

The equation of motion \motac\ is not consistent since all the
charges on the left hand side of \motac\ have the same sign. A
minimal modification that makes the equation consistent is to
write it as \eqn\neweqmot{ -\nabla^2 \varphi + \gamma \sum_i
\left( \delta(x-x_i) - { 2 \over v_6} \right) =0 } where $v_6$ is
the volume of the compact manifold. This term comes naturally from
the curvature of the four-dimensional spacetime, which, in the
approximation that we neglect the potential, is de Sitter space.
This positive curvature gives rise to a negative contribution to
the effective potential in the six internal dimensions.
It is reasonable to assume that the negative term is smeared over the
compact space as in this minimal modification, as long as the
transverse space is approximately homogeneous.\foot{
In compactifications with orientifold planes, there would also be
localized negative terms.  However, these would be cancelled by
the tensions of the branes which are present even
after brane/antibrane annihilation.
The extra energy of the inflationary brane/antibrane pair can be
expected to induce a smeared negative contribution over and above
the orientifold plane contribution.}

Note that this term does not arise for the Ramond fields since the
total charge is zero.

Let us now consider, for simplicity, the case of a single brane
and a single antibrane. In order to compute $\eta$ we compute the
Laplacian of the potential $V$ with respect to $x_1$. We get
\eqn\getlapl{
 \nabla^2_{x_1}  \varphi_{x_2}(x_1) = - {2\over v_6} \gamma
} The subscript in $\varphi$ indicates that this is the potential
due to the brane at $x_2$. For a pair of branes
the potential is $V = 2 +  \gamma
\varphi_{x_2}(x_1)$. The Laplacian has a constant negative value
\getlapl . We see that this implies that there exists at least one
direction in which the second derivative has a value $V'' \sim
\gamma \varphi'' \leq - { \gamma^2 \over 3 } { 1 \over v_6} $, since
there are six transverse dimensions. When we compute the
contribution to $\eta$ the factor $v_6$ cancels out.

When there are many fields, one should consider $\eta$ as a
matrix. In order to have slow roll inflation we need to demand
that the matrix has no negative eigenvalue that is too large. If
we have a large negative eigenvalue, then even if the scalar field
is not initially rolling in that direction, it will typically
start moving in this direction after a few e-foldings. The
discussion above implies that $\eta$, viewed as a matrix, has an
eigenvalue more negative than \eqn\boundeig{
 \eta|_{eigenvalue} \leq { - {2 \over 3} } \ .
}
This implies that at least one of the moduli acquires a
tachyonic mass $m^2 \leq -2H^2$, which typically prevents
a prolonged stage of inflation.

A similar analysis can be carried out for the general case of a
Dp-brane/anti-Dp-brane system. It is easy to see that the  only
change is that the coefficient ${2 \over 3}$ in \boundeig\ is
replaced by ${4 \over (9-p)}$. More interestingly, the above
analysis can  also be applied   to the case of Dp-branes at
angles.
 By this we mean a system of slightly misaligned  branes and
orientifold planes, \Garcia. The supersymmetry breaking scale in
such a system is controlled parametrically by an angle which
measures the relative orientation of the branes. For small values
of this angle, the vacuum energy, $V \sim \sum_iT_i$,  obtained by
summing over all the branes and planes,  can be  much  smaller
than the tension of any individual brane or plane. The force on a
brane in such a system  arises due to graviton-dilaton and RR
exchange. In these systems there can be a cancellation between the
graviton-dilaton and the RR force in such a way that the resulting
force, computed with non-compact  ``internal"  dimensions, is
parametrically smaller than the value of the cosmological
constant. Once the internal dimensions are compact, we have to
make some modification of the gravitational equation in order to
make it consistent. The simplest modification is to add a constant
term on the right hand side of the corresponding Laplace equation.
In this case the constant term will be proportional to the
effective four dimensional cosmological constant. Then, repeating
the analysis above, one finds that the resulting potential
satisfies the inequality
$$V^{''} \le -{\gamma^2 \over (9-p)} { 1\over v_6} T_p \sum_i T_i. $$
As a result, once again one obtains a value of $\eta$, \boundeig, with the coefficient
${2\over 3}$ replaced by ${4 \over (9-p)}$. In other words, both the potential and its
second derivative scale in
the same way with the small angle which supresses supersymmetry breaking, making $\eta$
independent  of this angle.

\appendix{B}{Computation of the $D3/\overline{D3}$
Potential in Warped Geometries}

To calculate the potential it is actually easier to turn things
around and view the D3-brane as perturbing the background and then
calculate the  resulting energy of the anti-D3-brane in this
perturbed geometry. This of course gives the same answer for the
potential energy of the brane-antibrane pair.

The coupling of the metric and the five-form to the D3-brane is
given by \dtact. On general grounds one expects that the changes
in the metric and $F_5$ caused by the D3-brane will vary in the
directions transverse to the brane. These directions are spanned
by the radial coordinate $r$ and the directions along $\rm X_{5}$.
It is useful to observe that the background can be written as
follows: \eqn\harmbac{ds^2 = h^{-{1 \over 2}} \left(-dt^2 +
{d\vec{x}}^2\right) + h^{{1\over 2}} \left(dr^2 + {r^2 \over R^2}
{\tilde g}_{ab} dy^ady^b\right)} \eqn\fluxbac{
(F_5)_{rtx^1x^2x3}=\partial_r h^{-1},} where ${\tilde g}_{ab}dy^a
dy^b$ is the line element on $\rm X_{5}$, and $h(r)$ is given by
\eqn\defharm{h(r)={R^4 \over r^4}.}

It is easy to check that $h(r)$ is a  harmonic function in a
six-dimensional space spanned by $r$ and the directions along $\rm
X_{5}$, with metric \eqn\sixmet{ds_6^2 = dr^2 + {r^2 \over
R^2}{\tilde g}_{ab} dy^ady^b.}

Adding one additional D3-brane at a radial location $r_1$ results
in a perturbed background which is of the form \harmbac, but with
a harmonic function  now given by \eqn\newharm{h(r) = {R^4 \over
r^4} +\delta h(r). } $\delta h$ solves the equation $\nabla_6^2
\delta h(r) = C \delta^6(\vec{r} -\vec{r_1})$ in the
six-dimensional space \sixmet.\foot{The constant $C$ is determined
by the tension of the D3-brane.}  For $r \ll r_1$ a simple
calculation shows  that \eqn\valdelh{\delta h(r)={R^4 \over N} {
1\over r_1^4}} independent of $r$ and the detailed metric on $\rm
X_{5}$. In \valdelh\ the coefficient $N$ arises because the
ambient background is supported by $N$ units of charge, whereas
the perturbation we are interested in arises due to a single
D3-brane. From \newharm\ the resulting harmonic function is
\eqn\resharm{h(r) = R^4\left({ {1 \over r^4} + {1\over N} {1 \over
r_1^4}}\right).}

To determine the potential we now couple this new background to
the anti-D3-brane. The anti-D3-brane is described by an action of
the form \dtact, except that, as was mentioned before, the sign of
the Chern-Simons term is reversed relative to the case of a
D3-brane.  We also remind the reader that the antibrane is located
at $r=r_0$; we will assume that $r_1 \gg r_0$. Combining all these
results, after a simple calculation one recovers the desired
potential \valpot.

This calculation of the potential is valid for one brane-antibrane
pair. For one brane and $p$ antibranes, to leading order, \valpot\
is simply multiplied by $p$. Corrections to this leading-order
potential are suppressed  for small $p$.

\appendix{C}{Warped Inflation }

In this appendix we discuss how inflation would look if one managed
to fix the overall volume modulus without giving a mass to the
brane motion.
We argued above that the  low energy dynamics of the system is
described by the action \fullact. The radial position of the
D3-brane, $r_1$, will play the role of the inflaton below. We
define a canonically normalized field \eqn\defcn{\phi= \sqrt{T_3}
r_1} and $\phi_0= \sqrt{T_3} r_0$. The effective action is then
given by \eqn\actcn{ S= \int d^4x \sqrt{-g}\left({{\cal{R}} \over
16 \pi G_N}  +
  {1 \over 2} g^{\mu\nu}\partial_\mu \phi
\partial_\nu \phi -   {4\pi^2\phi_0^4 \over N} \left({1- {1 \over N} {\phi_0^4 \over \phi^4}}\right)\right) }
We have assumed that there are no significant additional terms in
the effective action \actcn .

This inflaton potential is extremely flat: the first term in the
potential, which is independent of the inflaton, is larger than
the second term by a factor proportional to
$({r_1\over{r_0}})^4$. This factor can be interpreted as the
relative redshift between the brane location $r_1$ and the
antibrane location $r_0$; as we explained in \S4, this redshift is
exponentially sensitive to the parameters of the model:
\eqn\redagain{ r_0/R =e^{-{2 \pi K \over 3 g_s M}}} where $g_s$ is
the string coupling and $K,M$ are integers that specify fluxes
turned on in the compactification.

The slow-roll parameters can now be calculated in standard
fashion. We will use conventions where $8 \pi G_N = M_{Pl}^2$. One
finds that \eqn\para{\eqalign{\epsilon \equiv {M_{Pl}^2 \over 2}
\Bigl({V^{'} \over V}\Bigr)^2 & \simeq {8 \over N^2} M_{Pl}^2\
{\phi_0^8 \over \phi^{10}} \cr \eta \equiv M_{Pl}^2 {V^{''} \over
V} & \simeq -{20 \over N} M_{Pl}^2\ {\phi_0^4 \over \phi^6}.}}
Slow-roll requires that $|\eta| \ll 1, |\epsilon| \ll 1$. Of these
the condition on $\eta$ is more restrictive. It   can be met by
taking \eqn\bounda{\phi \gg \left({20 \over N} M_{Pl}^2
\phi_0^4\right)^{1/6}.}

The number of e-foldings is given by
\eqn\nefold{N_e={1 \over
M_{Pl}^2} \int {V \over V^{'}} d \phi \simeq {N \over 24}
 {1\over M_{Pl}^2} {\phi^6 \over \phi_0^4}.}

Requiring $N_e \sim 60$ can be achieved by taking $\phi$ to be
sufficiently large and is compatible with the bound \bounda.

Finally, the adiabatic  density perturbations  are given by
(\liddle, page 186) \eqn\dpert{\delta_H = {1\over \sqrt{75} \pi}
{1\over M_{Pl}^3}{V^{3/2} \over V^{'}} ={\sqrt N_e\over 2 \sqrt
{75}} {\phi^5\over \phi_0^2 M_{Pl}^3}.} This quantity should be
equal to $1.9\cdot 10^{-5}$ at $N_e \sim 60$, when the
perturbations responsible for the large scale structure of the
observable part of the universe are produced.

After some algebra, $\delta_H$ can be expressed in terms of $N_e$
as follows: \eqn\dptwo{\delta_H = C_1 N_e^{5/6}\left({T_3 \over
M_{Pl}^4}\right)^{1/3} \left({r_0 \over R}\right)^{4/3}. } $C_1$
is a constant which is somewhat model dependent; using \dpert\ and
\actcn, one has \eqn\valcone{C_1=
{{3^{1/3}2^{3/2}\over{5\pi}}}\left(N\over{T_3 R^4}\right)^{1/6}}
and after using \defr, \valten\ one finds that $C_1 =0.39$ for the
model of \S4.\foot{$C_1$ increases by a factor $\sqrt{p}$ when
there are $p$ antibranes. While making numerical estimates we set
$p=2$.}

The four-dimensional Planck scale ($M_{Pl}^{-2}\equiv  8 \pi G_N$)
is given by \eqn\valmpl{M_{Pl}^2={2 V_6 \over (2 \pi)^7
{\alpha^\prime}^4 g_s^2}} where $V_6$ is the volume of the
Calabi-Yau.  This formula is strictly applicable only to a
Kaluza-Klein compactification, not a warped compactification of
the kind considered here. However, the approximation is a good one
since the graviton zero mode has most of its support away from the
regions with large warping (where its wave function is
exponentially damped.) We may express the brane tension as
\eqn\ratmpt{ {T_3\over{M_{Pl}^4}}= {{(2\pi)}^{11}\over{4}}g_s^3
{{\alpha^\prime}^6\over{V_6^2}}}  This dimensionless ratio
evidently depends on the string coupling constant and the volume
of the six compact dimensions. The value $T_3/M_{Pl}^4 \sim
10^{-3}$ is quite reasonable: it corresponds to $g_s \sim 0.1$ and
a Calabi-Yau volume of a characteristic size $(V_6)^{1/6} \sim
5\sqrt{\alpha^\prime}$. Larger values of $V_6$ lead to smaller
values for $T_3/M_{Pl}^4$, which make it easier to meet the
density perturbation constraints.

More important, for present purposes, is the factor $({r_0 /
R})^{4/3}$, which has its origins in the redshift suppression of
the potential that was emphasized in the discussion above. By
taking  this factor to be small enough we see that the constraint
on $\delta_H$, \dpert, can be met. As an example, taking ${T_3 /
M_{Pl}^4} \sim 10^{-3}$ and $N_e =60$, we find that $\delta_H
\approx 1.91\cdot 10^{-5}$ for ${r_0 / R} = 2.5 \cdot 10^{-4}$.
This condition on $r_0/R$ can easily be met for reasonable values
of the flux integers $K,M$. Taking $g_s =0.1$, we get $r_0/R =2.5
\cdot 10^{-4}$, with $K/M \simeq 0.4$. The latter condition can be
achieved using moderate values of flux, e.g. $K = 8, M = 20$.

Now that we have ensured that the various constraints can be met
in our model, it is worth exploring the resulting inflationary
scenario a little more. The energy scale during inflation can be
expressed in terms of $\delta_H$. One finds from \dptwo, and using
the fact that the potential is well approximated by the first term
in \valpot, that \eqn\energysc{ {V \over M_{Pl}^4}= {{2
\delta_H^3} \over {C_1^3 N_e^{5/2}}}.} Taking $\delta_H=1.91 \cdot
10^{-5}$, $N_e=60$, $C_1=0.39$ and $M_{Pl}=2.4 \cdot 10^{18}$ GeV
one finds that the energy scale is \eqn\esctwo{\Lambda\equiv
V^{1/4}=1.3 \cdot 10^{14} \,  {\rm GeV} \, . } This is
considerably lower than the GUT scale $\sim 10^{16}$ GeV. This low
scale of inflation is a generic feature of the scenario.

Next, it is straightforward to see  that $\delta_H$ is given in
terms of $V$ and $\epsilon$ by
\eqn\relepa{\delta_H={ 1\over 5 \pi
\sqrt{6\epsilon}}\left({V\over M_{Pl}^4}\right)^{1/2}.} Solving
for $V$ from \energysc\ gives \eqn\relepb{\epsilon={\delta_H
\over{ 75 \pi^2 C_1^3 N_e^{5/2}}}.}
 Taking $\delta_H=1.91 \cdot
10^{-5}$, $C_1=0.39$,$N_e=60$ gives \eqn\valep{\epsilon=1.54 \cdot
10^{-11},} a very small number. The ratio of the anisotropy in the
microwave background generated by gravitational waves to that
generated by adiabatic density perturbations is given (at large
$l$) by \eqn\ratgrav{r \simeq 12.4 \epsilon.} In our model this is
very small, so the anisotropy is almost entirely due to density
perturbations.

Finally, $\eta$ can be related to $N_e$, and is given by
\eqn\rateta{\eta=-{5 \over 6} { 1\over N_e}.} Setting $N_e=60$
gives \eqn\valetab{\eta=-0.014.} Clearly, as we mentioned above,
$|\eta| \gg \epsilon$. The tilt parameter is given by
\eqn\valt{n=1-6 \epsilon + 2 \eta \simeq 1+2 \eta \approx 0.97,}
in excellent agreement with observational data from WMAP.

In summary, in our model the scale of inflation $\Lambda$ \esctwo\
is generically low.  Most of the anisotropy originates from
adiabatic density perturbations, since $\epsilon$ is extremely
small, and the tilt in the spectrum, \valt, is determined by
$\eta$. The values for these parameters are nearly
model-independent: they are almost entirely determined by the
observed value for $\delta_H$ and by the number of e-foldings,
$N_e$.

\appendix{D}{Eternal Inflation }

At large $\phi$, the potential $V(\phi)$ in \actcn\ becomes extremely flat.
For flat potentials, the force pushing the field $\phi$ down
becomes very small, whereas the amplitude of inflationary
fluctuations remains practically constant. As a result, the motion
of the field $\phi$  at large $\phi$ is mainly governed by quantum
jumps. This effect is known to lead to eternal inflation
\refs{\vilet,\linet}.

Eternal inflation leads to formation of a fractal structure of the
universe on a very large scale. It occurs for those values of the
field $\phi$ for which the post-inflationary amplitude of
perturbations of the metric $\delta_H$ would exceed unity \linde.  In
our case $\delta_H$ is proportional to $\phi^5$, cf. \dpert. Since
the amplitude of the density perturbations is $\delta_H \sim 1.9
\cdot 10^{-5}$ in the observable part of the universe, eternal
inflation should occur for all values of the field $\phi$ that are
greater than $10 \cdot \phi_{60}$. Here $\phi_{60}$ is the value
of the field at the moment starting from which the universe
inflated $e^{N_e} \sim e^{60}$ times. In other words, if $r_{60}$
is the brane separation corresponding to the moment when the
large-scale structure of the observable part of the universe was
produced, then the regime of eternal inflation occurred when the
brane separation was ten times greater than $r_{60}$. The
possibility of eternal inflation in our model is very interesting
since this regime makes the existence of inflation much more
plausible: even if the probability of initial conditions for
eternal inflation is small, the universes (or the parts of the
universe) where these conditions are satisfied rapidly acquire
indefinitely large (and ever growing) volume \lindeXX.

\appendix{E}{Exit from Inflation }

In this appendix we comment on the exit from inflation through
brane-antibrane annihilation.

 The brane-antibrane potential used in our analysis of
inflation is no longer valid when the brane separation is
comparable to the string length.  At that stage a tachyon appears
and then condenses.  (In this sense, our model, like all the brane
inflation models described in \quevedo, is a particular version of
the hybrid inflation scenario \hybrid.)  One may attempt to use
the properties of this brane-antibrane tachyon \refs{\Sen,\tachac}
to describe the exit from inflation.
Here we will show that one of the possible problems of this scenario, the overproduction of cosmic strings \refs{\KofLinde,\tyecosmic}, is ameliorated by the warped geometry.

In the case of a merging brane-antibrane pair, the tachyon is a
complex field and there is a $U(1)$ symmetry.  Formation of cosmic
strings associated with the $U(1)$ symmetry breaking leads to
large-scale perturbations of the metric which are compatible with the
current observations of the cosmic microwave anisotropy
\BennettBZ\ only if $G_N\, T_1= {T_1 \over 8\pi M_{Pl}^{2}}\la 10^{-7}$,
where $T_1$ is the cosmic string tension \liddlenew. This tension
can be evaluated either by the methods of \senvertex, or  by
identifying cosmic strings with D1-branes. In the usual case (i.e.
ignoring warping) one has
 \eqn\tension{ T_1 =
 {1 \over {2\pi g_s  \alpha'}}.}
The requirement $G_N\, T_1={T_1 \over 8\pi M_{Pl}^{2}}\la 10^{-7}$ reads
 \eqn\T{G_N \, T_1=  {g_s
\over 16\pi} {(2 \pi l_s)^6\over V_6}  \la 10^{-7} , }
i.e.
 \eqn\VV{V_6 \ga 2 \times 10^5\ g_s\ {(2 \pi l_s)^6} . }
This shows that the cosmic string contribution to the
perturbations of the metric produced after inflation is
unacceptably large unless the volume of the compactified space
$V_6$ is at least five orders of magnitude greater than  $g_s
(2 \pi l_s)^6$.

In the brane inflation models of \S3, \S4, however, the relevant tension is redshifted by the warped geometry, which leads to exponential suppression of $T_1$:
\eqn\T{ T_1 =
 {1 \over {2\pi g_s  \alpha'}} e^{-{4 \pi K \over 3 g_s
M}}.} As a result, the undesirable contribution of cosmic strings
(D1-branes) to perturbations of the metric becomes exponentially
suppressed.

\appendix{F}{Fine-tuning of the Potential when the Superpotential Depends on the Inflaton Field}

In this appendix we study a toy model in order to make more
precise our statements concerning the degree of fine-tuning which
is required for slow-roll brane inflation.  We should note here
that we will be discussing the degree to which the inflaton potential
itself must be tuned.  In a given string model, one cannot directly
tune the potential, but only vary choices of the background data
like fluxes, compactification manifold, or brane positions.  It
could be that the tuning required in terms of this data is more or
less severe than our estimate below, but explicit string
calculations of
the relevant superpotentials will be necessary to determine this.

Before studying the example, let us mention how small the inflaton
mass term must be for a given model of slow-roll inflation to be
compatible with experiment. The goal is to have a long stage of
inflation producing metric fluctuations with a fairly flat
spectrum.  Recent observations suggest that, modulo some
uncertainties, the tilt is $n_s \approx 1 + {2m_\phi^2\over 3H^2}
= 0.97 \pm 0.03$ \refs{\BennettBZ,\ContaldiHI}.  This is
compatible with an inflaton mass $|m^2_\phi|/H^2 \sim 10^{-1}
-10^{-2} $.

This could be achieved through fine-tuning of $m^2_\phi$ by only
about one part in 100. Thus, the fine-tuning that we need to
perform is not extraordinary.  Given the large number of possible
compactifications, the existence of some configurations which
allow inflation seems quite likely.

We now turn to an example which illustrates this point.  Consider
a D3-brane transverse to a warped compactification; we would like
to know how the (brane-antibrane) inflaton mass terms vary as the
inflaton-dependence of the superpotential varies.

The K\"ahler potential for the volume modulus  and the D3-brane
field $\phi$ takes the form $K(\rho,\bar \rho,\phi,\bar\phi) = - 3
\log \left(\rho + \bar \rho - k(\phi,\bar\phi)\right)$.  We will
work in the vicinity of the point $\phi=\bar \phi = 0$ in moduli
space, where $k(\phi,\bar\phi)=\phi\bar \phi$. We choose a
superpotential  of the form
 \eqn\wnpis{ W(\rho, \phi) =W_{0} + g(\rho)f(\phi) } where
$g(\rho)$ is an arbitrary function of $\rho$, $f(\phi)=(1+\delta\,
\phi^2)$, and $W_{0}$ and $\delta$ are constants.  This is a
slight generalization of the superpotential in \KKLT, which
corresponds to $\delta=0$ and $g(\rho)= Ae^{-a\rho}$.

One can now calculate the supergravity potential $V^F = e^K
(g^{i\bar j} D_i W \overline{D_j W} - 3 |W|^2) $ for the two
complex fields $\rho, \phi$. The exact potential has a simple
dependence on $\rm Im \,\rho$ and $\rm Im\, \phi$ which shows that
the point $\rm Im\, \rho = \rm Im\, \phi=0$ is an extremum of the
potential (it is a minimum, at least for small $\phi$). Therefore
we will present here the exact potential $V^F (\sigma, \psi)$ as a
function of $\rm Re\, \rho=\sigma$ and $\rm Re\, \phi=\psi$ at
$\rm Im \,\rho= \rm Im\, \phi=0$. \eqn\potLiam{V^F(\sigma, \psi)
={{1\over 6(\sigma- \psi^2/2)^2}\left({ 2 \delta^2 \psi^2
g(\sigma)^2+ f(\psi) g'(\sigma)g(\sigma)\left({-
{3W_0\over{g(\sigma)}}+ \sigma f(\psi)-f(\psi)-2}\right)}\right)}}
We are interested in the total potential $V^F(\sigma, \psi)+
V_{\overline D3}$ at small $\psi$, where $V_{\overline D3}$ is the
potential due to the antibrane (cf. \nowvis). We may therefore use
the stabilization of the volume in the first approximation  at
$\psi^2=0$ and calculate the potential at the AdS critical point
$\sigma_{c}= r_{c}$, where, using $D_\rho W|_{\phi=0}=0$, one
finds \eqn\AdS{W_0=-g(\sigma_{c})+ {2\over 3} \sigma_{c}
g'(\sigma_{c}) \ , \qquad V_{AdS}=-{(g'(\sigma_{c}))^2\over
6\sigma_{c}}} We now change variables to $\psi^2= {2\over 3}
\sigma_{c}\varphi^2$, where $\varphi$ is a  field with the
canonical kinetic term $(\partial \varphi)^2$.  We find \eqn\Liam
{V^F (\sigma_{c}, \varphi)= {1\over 6 \sigma_{c}(1-
\varphi^2/3)^2}\left({-(g')^2+ { 4 \delta^2 g^2\over 3 } \varphi^2
-  { 2\delta g g'\over 3 } \varphi^2 (1+ {2 \delta\over 3}
\sigma_{c}\varphi^2) +{4\over 9} \delta^2 \varphi^4\sigma_{c}^2
}\right)}
 From the antibrane we get the additional contribution mentioned above.
 Keeping terms up to those quadratic in $\phi$, we finally arrive
 at
\eqn\answer {V^F (\sigma_{c}, \varphi)+ V_{\bar D3}(\sigma_{c},
\varphi) \approx V_{dS} +{2 \varphi^2 \over 3}   \left (V_{dS} + {
1\over 6 \sigma_{c} }(2 \delta^2 g^2 -  \delta g' g )\right)} Here
$V_{dS}$ is the value of the potential at the de Sitter minimum,
\eqn\dS{V_{dS}= V_{AdS}+{ D\over 4\sigma^2_{c} }=
-{(g'(\sigma_{c}))^2\over 6\sigma_{c}}+{ D\over
4\sigma^2_{c}}\equiv 3H^2 } The mass-squared of the field $\phi$
is \eqn\cond{m^2_\phi =2H^2+ {2 |V_{AdS}|\over
3}\left[2\left(\delta {g\over g'}\right)^2-\delta {g\over g'}
\right]} To make $m^2_\phi$ small, we need $ \delta {g\over g'} >
0$  as well as ${ |V_{AdS}|\over 3}\left[2\left(\delta {g\over
g'}\right)^2-\delta {g\over g'} \right]\approx -H^2$. If the
parameters of the model were arbitrary then this would certainly
be possible.

We will express our results in terms of a parameter $\beta =
\delta {g\over g'}$:
\eqn\condrr{m^2_\phi =2H^2- {2\over 3}
|V_{AdS}|(\beta - 2\beta^2)= 2H^2\left(1- {
|V_{AdS}|\over V_{dS}}(\beta - 2\beta^2)\right)}
For $\beta =0$ we
recover the ``conformal'' result \eqn\cond{m^2_\phi =2H^2}

As a simple example, if $g(\rho)= A e^{-a\rho}$, as in \KKLT, we
find $\beta =- {\delta\over a}$.   However, let us assume, as in
\KKLT, that $|V_{AdS}|\gg V_{dS}$. Then for the simple value
$\beta = 1$ (i.e. $\delta = - a$) we have
\eqn\condcc{m^2_\phi
=2H^2\left(1+ {|V_{AdS}|\over V_{dS}}\right) \approx {2\over 3}
|V_{AdS}| \gg 2H^2}

Thus, whereas it is true that our knowledge of $W(\rho,\phi)$ is
not particularly good, our absence of knowledge does not allow us
to say much about $m^2_\phi$. The only thing we can say is that in
our particular example, for $|V_{AdS}|\gg V_{dS}$, this mass can
be fine-tuned to take almost any value.\foot{Incidentally, Eq.
\condcc\ implies that if one does not make any fine-tuning, then
for the model described in \KKLT, with $V_{dS} \sim 10^{-120}$ in
Planck units, the typical mass squared of the D3 brane moduli
fields is expected to be ${\cal{O}}(|V_{AdS}|)$, which can be
extremely large. This result may have interesting phenomenological
implications. } In particular, one has a flat potential with
$m^2_\phi=0$ for \eqn\condbb{\beta ={1\over 4}\left(1\pm
\sqrt{1-{8 V_{dS}\over |V_{AdS}|}}\right)} This equation always
has solutions for $|V_{AdS}|\geq 8 V_{dS}$.  For $|V_{AdS}|\gg 8
V_{dS}$, the solutions are: \eqn\condaa{\beta_1={\delta_1\over a}
={1\over 2}-{V_{dS}\over |V_{AdS}|} \approx {1\over 2}} and
\eqn\cond{\beta_2={\delta_2\over a} ={V_{dS}\over |V_{AdS}|} \ll
1}

In order to satisfy one of these two conditions and have
$m^2_\phi=0$ one can fine-tune either the ratio
$V_{dS}\over{|V_{AdS}|}$ (as was done in \KKLT) or the coefficient
$\delta$ in the superpotential. In order to prove that inflation
in this scenario is impossible, one would need to prove that
neither of these types of fine-tuning is possible.

It is instructive to compare this situation with the problem of
realizing the chaotic inflation scenario in ${\cal{N}}=1$
supergravity. Let us consider a canonical K\"ahler potential
$K=\bar \phi \phi + \bar \sigma_i \sigma_i$, where $\phi$ is the
inflaton field and $\sigma_i$ are some other fields. If the
superpotential is a function of the fields $\sigma_i$ but not of
the field $\phi$, then the potential of the scalar fields has the
general structure as a function of the real part of the field
$\phi$, \ $V = e^{\phi^2}~ V(\sigma_i)$, which implies that
$m^2_\phi =3H^2$, i.e. $\eta =1$.

One can resolve this problem by introducing a superpotential
depending on the inflaton field, just as we did in this appendix.
However, in the simplest version of chaotic inflation one needs
the inflaton field to be at $\phi \gg 1$, in Planck mass units,
and to change significantly, by $\Delta \phi = {\cal{O}}(1)$,
during the last 60 e-folds. It is this last part that causes
substantial difficulties for inflation in ${\cal{N}}=1$
supergravity. It is always possible to find a superpotential which
depends on the inflaton field $\phi$ in such a way that the
potential becomes flat in the vicinity of one particular point.
However, one must do this for all $\phi$ in a large interval
$\Delta \phi = {\cal{O}}(1)$. One needs enormous {\it functional
fine-tuning} in a large interval at $\phi \gg 1$, where the term
$\sim e^{\phi^2}$ grows very fast.

Meanwhile, in our case the situation is much better. Instead of
 a functional fine-tuning in a large interval of $\phi$
we need to make a fine-tuning at a single
point $\sigma = \sigma_{c}$, $\phi = 0$. In
order to estimate the required degree of fine-tuning, let
us e.g. fix $\beta = 1/2$ and change the ratio $|V_{AdS}|/V_{dS}$ in
Eq. \condrr\ in the interval
$0<|V_{AdS}|/V_{dS}<4$. As one can easily see, in this case
the mass squared of the inflaton field
changes from $2H^2$ to $-2H^2$. In approximately 1\% of this
interval the condition $n_s \approx 1
+ {2m_\phi^2\over 3H^2}
= 0.97 \pm 0.03$ is satisfied. On the other hand, if
this condition is substantially violated, which
 happens in the main part of this interval, then inflation
becomes either too short or impossible,
and the universe most probably becomes unsuitable for life.

Finally, if inflation can be eternal (and it can be eternal in the
models of \S3,4, see Appendix D), then the parts of
the universe where eternal inflation is possible
have an
indefinitely large and ever-increasing volume.  For this reason,
regions of the universe where eternal inflation does occur, however
improbable that may have been, are in some sense favored.  One
could therefore argue that the problem of
fine-tuning in inflationary cosmology
is not as  dangerous as one could expect, and sometimes it may not even be
particularly relevant.

\listrefs

\bye

\end